\documentclass[]{aastex631}

\usepackage{amsmath}
\usepackage{amsfonts}
\usepackage{graphics}
\usepackage{color}
\shorttitle{TDE in AT2019dsg}
\shortauthors{Mohan et al.}
\graphicspath{{./}{figures/}}
\begin{document}
 
\title{High-resolution VLBI observations of and modelling the radio emission from the TDE AT2019dsg}

\correspondingauthor{Prashanth Mohan}
\email{pmohan@shao.ac.cn}

\correspondingauthor{Tao An}
\email{antao@shao.ac.cn}

\author{Prashanth Mohan}
\affiliation{Shanghai Astronomical Observatory, \\ Key Laboratory of Radio Astronomy, Chinese Academy of Sciences, \\Nandan Road 80, 200030, China}

\author{Tao An}
\affiliation{Shanghai Astronomical Observatory, \\ Key Laboratory of Radio Astronomy, Chinese Academy of Sciences, \\Nandan Road 80, 200030, China}

\author{Yingkang Zhang}
\affiliation{Shanghai Astronomical Observatory, \\ Key Laboratory of Radio Astronomy, Chinese Academy of Sciences, \\Nandan Road 80, 200030, China}

\author{Jun Yang}
\affiliation{Department of Space, Earth and Environment,\\ Chalmers University of Technology, Onsala Space Observatory,\\ SE-439 92 Onsala, Sweden}

\author{Xiaolong Yang} 
\affiliation{Shanghai Astronomical Observatory, \\ Key Laboratory of Radio Astronomy, Chinese Academy of Sciences, \\Nandan Road 80, 200030, China}

\author{Ailing Wang}
\affiliation{Shanghai Astronomical Observatory, \\ Key Laboratory of Radio Astronomy, Chinese Academy of Sciences, \\Nandan Road 80, 200030, China}
\affiliation{University of Chinese Academy of Sciences,\\ 19A Yuquanlu, Beijing 100049, China} 	


\begin{abstract}
A tidal disruption event (TDE) involves the shredding of a star in the proximity of a supermassive black hole (SMBH). The nearby ($\approx$230 Mpc) relatively radio-quiet, thermal emission dominated source AT2019dsg is the first TDE with a potential neutrino association. The origin of non-thermal emission remains inconclusive; possibilities include a relativistic jet or a sub-relativistic outflow. Distinguishing between them can address neutrino production mechanisms. High-resolution very long baseline interferometry 5-GHz observations provide a proper motion of 0.94 $\pm$ 0.65 mas yr$^{-1}$ ($3.2 \pm 2.2~c$; $1-\sigma$). Modelling the radio emission favors an origin from the interaction between a decelerating outflow (velocity $\approx$ 0.1 $c$) and a dense circum-nuclear medium. The transition of the synchrotron self-absorption frequency through the observation band marks a peak flux density of 1.19 $\pm$ 0.18 mJy at 152.8 $\pm$ 16.2 days. An equipartition analysis indicates an emission region distance of $\geqslant$ 4.7 $\times$ 10$^{16}$ cm, magnetic field strength $\geqslant$ 0.17 G, and number density $\geqslant$ 5.7 $\times$ 10$^{3}$ cm$^{-3}$. The disruption involves a $\approx$ 2 $M_\odot$ star with a penetration factor $\approx 1$ and a total energy output of $\leqslant$ 1.5 $\times$ 10$^{52}$ erg. The outflow is radiatively driven by accretion of stellar debris onto the SMBH. Neutrino production is likely related to the acceleration of protons to PeV energies and the availability of a suitable cross-section at the outflow base. The present study thus helps exclude jet-related origins for non-thermal emission and neutrino production, and constrains non-jetted scenarios.
\end{abstract}

\section{Introduction}
When a star passes within the tidal radius of a galactic supermassive black hole (SMBH of mass $M_\bullet = 10^6 - 10^8$ $M_\odot$, where $M_\odot$ is the solar mass), its destruction by tidal forces leads to the fallback accretion of stellar debris \citep{1988Natur.333..523R,1989ApJ...346L..13E}. This usually powers a thermal flare in optical, ultra-violet (UV) and X-ray bands, with a luminosity comparable to bright Supernovae at $\approx 10^{43} - 10^{44}$ erg s$^{-1}$ \cite[e.g.][]{2020SSRv..216...85S}. The optical/UV luminosity generally declines with a power law index of $-5/3$ \cite[e.g.][]{2020SSRv..216...85S} over months to years, indicative of fallback accretion \citep{1989ApJ...346L..13E,1989IAUS..136..543P}. However, there may be small or significant deviations from this trend \cite[e.g.][]{2021ApJ...908....4V}. Further, 
the later phase (months to years) decline in the X-ray and UV bands may be more shallow (with an index of $-5/12$) and is indicative the prevalence of disk emission \citep{2011MNRAS.410..359L}. Outflows can be launched as a result of stream-stream collisions involving the interaction between the orbiting stellar debris streams and material accreting onto the SMBH \cite[e.g.][]{2016ApJ...830..125J,2020MNRAS.492..686L}. These can also originate as radiation pressure driven disk winds \citep{2009MNRAS.400.2070S}. 
Upon interaction with the surrounding circum-nuclear medium (CNM), the outflow produces a non-thermal, pre-dominantly synchrotron afterglow emission \cite[e.g.][]{2020SSRv..216..114R,2016ApJ...819L..25A}. Multi-wavelength observations of the afterglow can help identify the onset of this interaction and constrain the energy output from the TDE and the properties of the CNM \cite[e.g.][]{2020SSRv..216...81A}.

The transient AT2019dsg was discovered on 2019 April 9 by the Zwicky Transient Factory \citep{2019TNSTR.615....1N} and identified as a TDE in a host galaxy at a redshift of $z = 0.051$ \citep{2019ATel12752....1N, 2021ApJ...908....4V}, making it a rarely-seen nearby TDE.  \cite{2021NatAs...5..510S} reported a putative association of AT2019dsg with the 0.2 peta electron volt (PeV) energy neutrino event IC191001A, which was detected by the IceCube neutrino observatory on 2019 October 1 \cite[175 days after the TDE discovery; ][]{2019GCN.25913....1I} and is expected to be of astrophysical origin. A multi-wavelength analysis in the same study \citep{2021NatAs...5..510S} indicates a highly luminous TDE (optical peak luminosity of $\approx 3.5 \times 10^{44}$ erg s$^{-1}$) and suggests the presence of an active central engine embedded in an optically thick photosphere, which powers sub-relativistic outflows and provides a suitable site for the neutrino production. Optical polarimetric observations \citep{2020ApJ...892L...1L} show a decrease in linear polarization degree from 9.6\% near the optical peak around 2019 May 17 to 2.7\% on 2019 June 20 during the declining phase; this variation is attributed to a non-isotropic accretion disk or a relativistic jet. A concordance scenario \citep{2021NatAs...5..472W} proposes the presence of a relativistic jet that participates in the acceleration of protons and provides the necessary site for neutrino production. Alternatively, an off-axis jet may host sites for hadronic interactions and  neutrino production \citep{2020PhRvD.102h3028L}. An analysis of multi-wavelength radio observations of AT2019dsg \citep{2021arXiv210306299C} suggests that the emission originates from the interaction of a non-relativistic outflow with the surrounding medium. Based on the relatively low outflow velocity and energy output in the context of TDEs and Type Ib/c Supernovae, the study of \cite{2021arXiv210306299C} suggests that the neutrino association is less likely. \cite{2021arXiv210902648M} probe the nature of the radio emission through the associated synchrotron self-absorption, and use calculations based on an equipartition model to infer the presence of a freely expanding or decelerating non-relativistic outflow. The complex observational signatures and associated inferences are thus inconclusive as to the origin of the non-thermal emission, which may originate from either a relativistic jet or a non-relativistic outflow. Clearly distinguishing between them is crucial to constrain the TDE properties (physical and geometric properties of the black hole--disk system and the disrupted stellar system), the nuclear environment of the host galaxy, the central engine activity and to clarify the neutrino production mechanisms.

Very long baseline interferometry (VLBI) observations at milli-arcsecond (mas) resolutions can help measure the size of the radio afterglow, constrain the proper motion of the emitting component \cite[e.g.][]{2016MNRAS.462L..66Y,2018Sci...361..482M,2020ApJ...888L..24M}, and provide inputs to discriminate the TDE and CNM properties. Of those TDEs that are detected in radio \cite[e.g.][]{2020SSRv..216...81A,2021NatAs...5..491H}, 
only a few have been imaged using VLBI, with large differences in radio luminosity evolution, environmental properties, and the level of collimation of the outflow. Besides, extensive monitoring campaigns have only been conducted on \textit{Swift} J1644$+$5734 \cite[e.g.][]{2011Sci...333..199L,2011Natur.476..425Z, 2012ApJ...748...36B,2013ApJ...767..152Z,2016MNRAS.462L..66Y,2018ApJ...854...86E,2021ApJ...908..125C} and Arp 299-B AT1 \cite[e.g.][]{2018Sci...361..482M} owing to the presence of a relativistic jet,  ASASSN-15oi \citep{2021NatAs...5..491H} due to a peculiarly delayed radio flaring (over months to years timescales), ASASSN-14li \cite[e.g.][]{2016Sci...351...62V,2016ApJ...819L..25A,2016ApJ...832L..10R,2018MNRAS.475.4011B} being relatively close proximity ($z =$ 0.02), and AT2019dsg owing to its persistent radio emission and potential neutrino association \citep{2021NatAs...5..510S,2021MNRAS.504..792C,2021arXiv210306299C,2021arXiv210902648M}. A multi-wavelength observation campaign was carried out to study AT2019dsg and its host galaxy \citep{2021MNRAS.504..792C}. The results include the inference of a central SMBH of $5.4 \times 10^6 \, M_\odot$, derived from the optical spectroscopy. In addition, they report the detection of a compact source with no significant displacement (one sigma astrometric uncertainty of $\approx$ 6 mas) from the phase center, as inferred from the 5 and 1.4-GHz radio monitoring observations made by the electronic Multi-Element Remotely Linked Interferometer Network (\textit{e}-MERLIN) up to 180 days after the TDE.  

\section{Observations}

We employ the VLBI technique to monitor any emission structure change of AT2019dsg at high resolutions, motivated by potentially identifying a (collimated) relativistic jet and determining the source size. This makes it one of only three thermal emission-dominated TDEs with VLBI monitoring, the others being ASASSN-14li \citep{2016ApJ...819L..25A} and CNSS J0019+00 \citep{2020ApJ...903..116A}. Our observations were carried out with the European VLBI Network (EVN) at 5 GHz spanning three epochs at 200, 216 and 304 days after the TDE, covering the declining phase of the source radiative evolution. As the source was successfully detected in all three epochs, this provides unique constraint information: flux densities during the declining phase, and kinematics including source size and astrometry to infer any significant proper motion. The first and third observations were made in the standard disk-recording mode, involving a maximum of 20 telescopes. The second observation was made in the electronic-VLBI mode (\textit{e}-EVN). Details of the observation logs are presented in Table \ref{tab:obs} and the data reduction is presented in Appendix \ref{VLBIobs}. 

\section{Results and Discussion}

\begin{figure}
    \centering
    \includegraphics[scale=0.47]{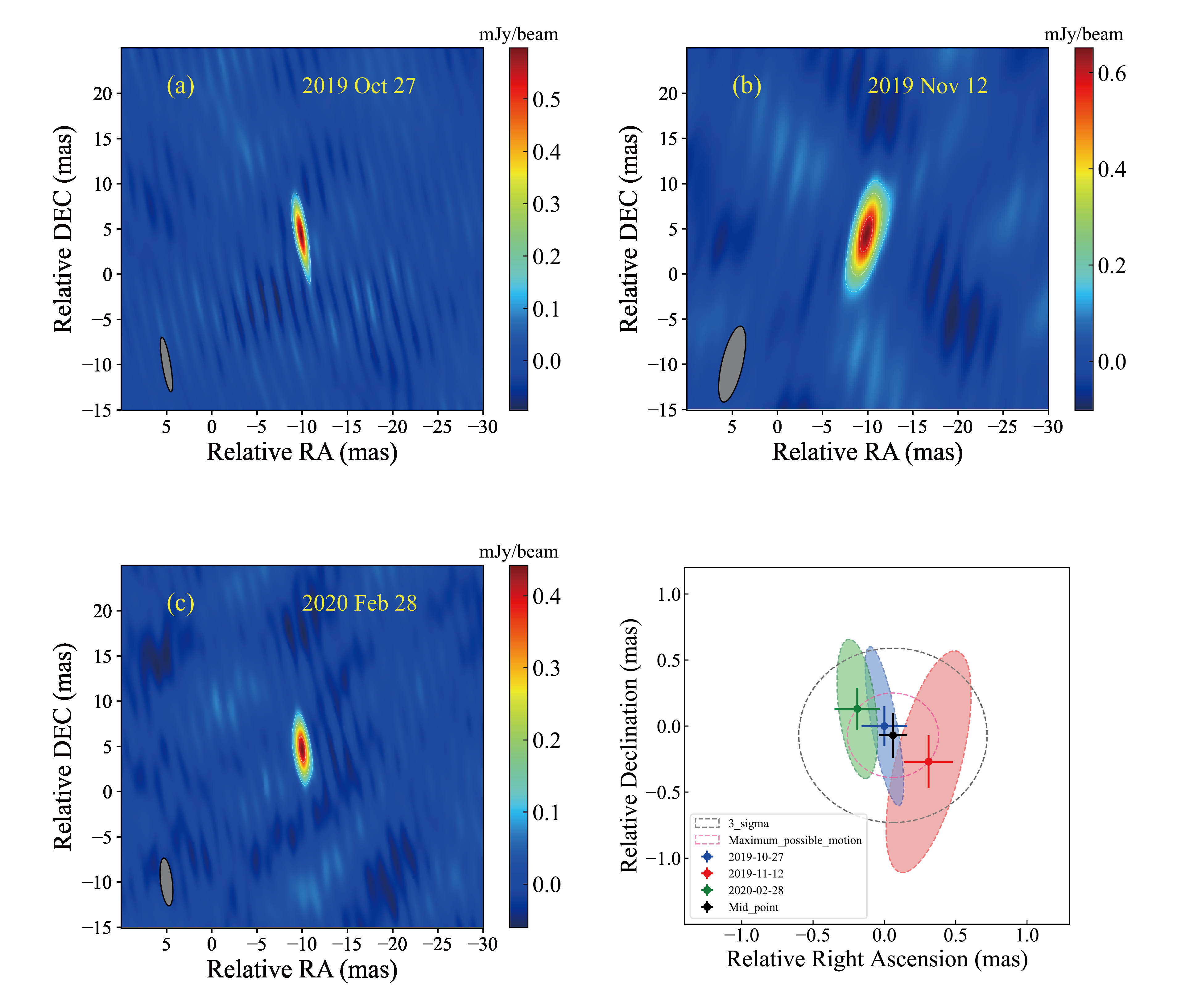}
    \caption{Top row and bottom left: high-resolution images from the three EVN observations showing a compact, unresolved emitting source. Bottom right: constraints on the source peak positions from astrometric measurements, relative to position of calibrator source CAL2. The mid-point is at $(0.06~{\rm mas},-0.07~{\rm mas})$ and the radius of the inner red dashed circle is the maximum relative position change $d = 0.32$ mas. The $3-\sigma$ error on $d$ is the outer black dashed circle which mostly encloses the source positions indicating that there is no significant shift in the emission center across the three observation epochs. The image center is at right ascension (RA) $\alpha =$ 20$^h$57$^m$ 02$^s$.965 and declination (Dec) $\delta = +14^\circ 12^\prime 16^{\prime\prime}.290$ (J2000); the colour bar denotes the brightness scale. The imaging parameters and position measurements of each epoch are presented in Tables \ref{tab:obs} and  \ref{tab:err}. The peak positions are measured using the reference from the phase-referencing calibrator J2052+1619, with positional astrometric uncertainties of 0.14 mas and 0.20 mas along the RA and Dec directions, respectively. The variation of the peak position falls within one fifth of the restoring beam marked by the ellipse in the bottom-left corner of each panel.}
    \label{tar-img}
\end{figure}

Figure~\ref{tar-img} shows the EVN images of AT2019dsg during the three epochs and the constraints on the peak emission positions with respect to the phase center. The astrometric parameters involving the calibrators and target source including errors are presented in Table \ref{tab:err}. 
With a maximum expansion distance of $0.32 \pm 0.22$ mas ($1-\sigma$) that spans 123.7 days, the proper motion is $0.94 \pm 0.65$ mas yr$^{-1}$ (3.2 $\pm$ 2.2 $c$)\footnote{Assuming a flat cosmology with parameters H$_{0}$ = 70 km s$^{-1}$ Mpc$^{-1}$, $\Omega_{m} = 0.27$, $\Omega_{\Lambda} = 0.73$, at $z = 0.051$, an angular separation of 1 mas corresponds to a projected size of $\approx$ 1 pc \citep{2006PASP..118.1711W}, and a proper motion of 1 mas yr$^{-1}$ corresponds to an expansion speed of 3.4 $c$}. At the $1-\sigma$ level of uncertainty, we cannot rule out a relativistic expansion. However, as the proper motion $\lesssim 1.4-\sigma$, it may be considered to be statistically insignificant (with a requirement to be $\geqslant 3-\sigma$).
A discernible proper motion may be expected in the case of an off-axis jet. For an on-axis view, Doppler beaming of emission along the observer's line of sight is expected to result in high brightness temperatures \cite[$\geqslant 10^{10}$ K, the case for blazars and jetted active galactic nuclei, e.g.][]{2020ApJS..247...57C}. Neither of these scenarios are applicable here owing to the absence of a statistically significant proper motion (large uncertainties at the $3-\sigma$ level) and brightness temperatures of  $\sim 10^9$ K and $\geq 10^8$ K during the first and third observation epochs (see Table \ref{tab:parm}). These and the detection of an unresolved compact structure indicate the presence of a non-thermal component, and are consistent with afterglow emission from the interaction of an outflow with the surrounding CNM, which produces an expanding forward shock.

\begin{figure}
\centering
\includegraphics[scale=0.4]{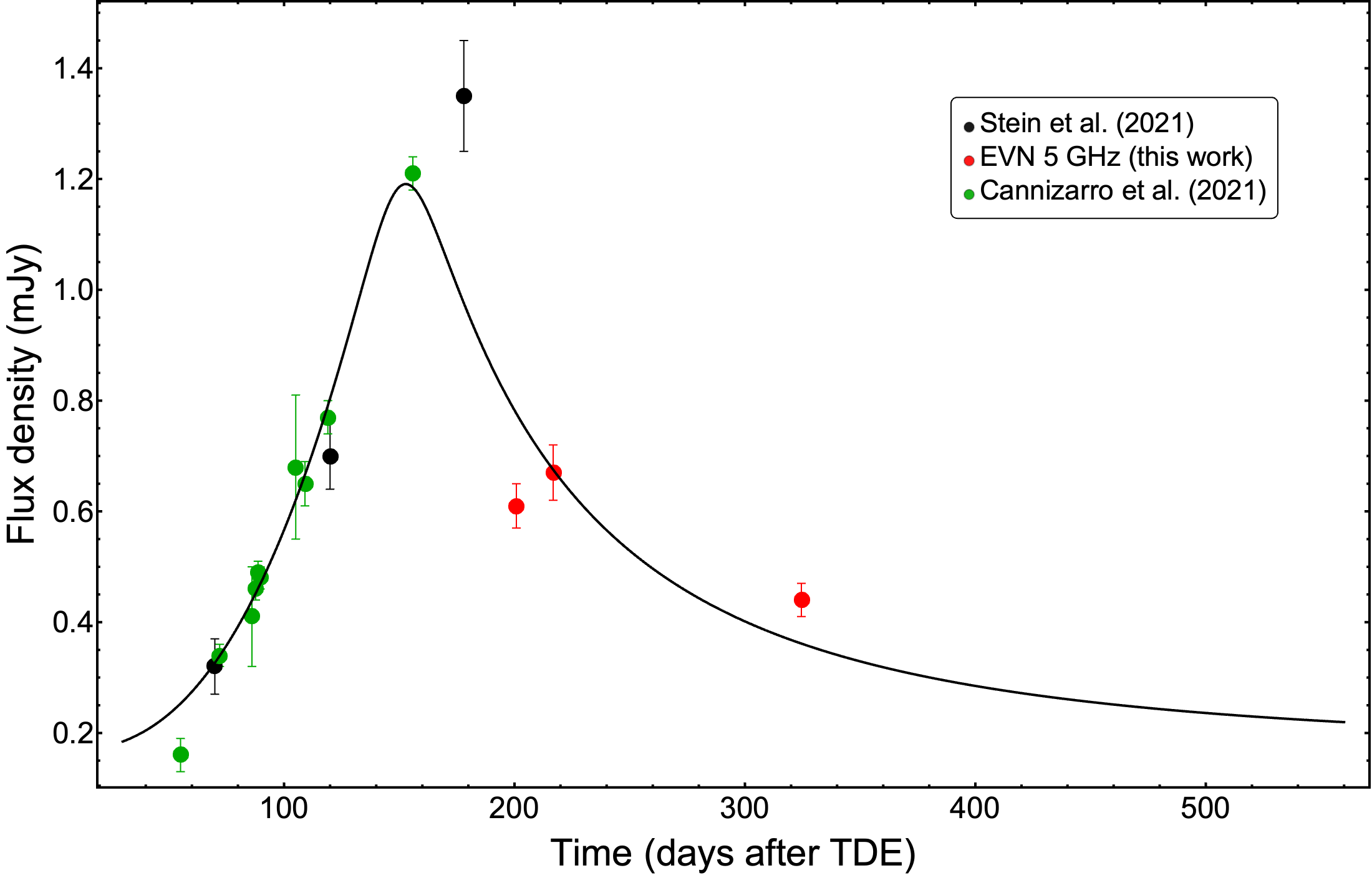}
\caption{Compiled 5 GHz light curve fit with a smooth broken power-law plus a linear trend (see Appendix \ref{lcfitting}). The black and green data points cover the rising phase and peak and are sourced from \cite{2021NatAs...5..510S} and \cite{2021MNRAS.504..792C}. The red data points during the declining phase are from our EVN observations. The fit yields a rising slope $\alpha_1$ of $2.38 \pm 0.93$, a declining slope $\alpha_2$ of $-2.56 \pm 1.31$, peak flux density $F_{\nu_p} = 1.19 \pm 0.18$ mJy at a time $t_p = 152.80 \pm 16.19$ days.}
\label{lcurvefit}
\end{figure}

The compiled 5-GHz light curve consists of flux densities reported in \cite{2021NatAs...5..510S} and \cite{2021MNRAS.504..792C}, as well as the measurements from our EVN observations. A smooth broken power law is used to fit the light curve (see Appendix \ref{lcfitting} and Table \ref{tab:fitpars} for the fit parameter estimates), presented in Figure \ref{lcurvefit}. The peak flux density $F_{\nu_p}$ is inferred to be 1.18 $\pm$ 0.18 mJy at a time $t_p$ of 152.8 $\pm$ 16.2 days. The power-law indices during the rising and declining phases are $\alpha_1 = $ 2.4 $\pm$ 0.9 and $\alpha_2 =$ $-$2.6 $\pm$ 1.3, respectively. The peak flux density, associated time and the power law temporal indices are used as inputs in the modeling and interpretation. For a spherically expanding forward shock with a constant electron injection energy index $p$, a time dependence of size $R \propto t^\alpha$, of magnetic field strength $B \propto R^{-2} \propto t^{-2 \alpha}$ and the proportionality constant (for the number of electrons per unit energy within the region) $N_0 \propto B^{2} \propto t^{-4 \alpha}$ result in optically thick and thin flux densities that scale as $\propto t^{3 \alpha}$ and $\propto t^{-(2+p) \alpha}$ respectively. The expansion velocity $v \propto t^{\alpha-1}$; comparing the measured maximum speed of 0.57 $c$ with that inferred from the VLBI measured size of 0.17 mas (0.17 pc) and the 5 GHz flux density peak time $t_p$ of 152.4 days, $\alpha \geq$ 0.59, with the lower limit implying a deceleration of the expanding shock. This is consistent with the decelerating solution (index of 0.63) based on an equipartition analysis as inferred in \cite{2021arXiv210902648M} for a radio-emitting outflow interacting with the surrounding CNM. Using $\alpha = 0.59$, and $p = 2.7 \pm 0.2$ inferred in \citet{2021arXiv210306299C} under the assumption of a spherically expanding outflow, the optically thin flux density scales as $\propto t^{-2.8}$, consistent with the measured $\alpha_2 = -2.6 \pm 0.3$. Hence, we adopt these as physical parameters for subsequent calculations in the radiative models. The optically thin spectral index $\alpha_s$ depends on the observation frequency as $\propto \nu^{-(p-1)/2} \propto \nu^{-0.85}$, indicating that the emission is dominated by a steep-spectrum component. 
The radio luminosity at 5 GHz is then $\nu L_\nu$ $\approx$ $4 \pi D^2_L F_{\nu_p} (1+z)^{-(1-\alpha_s)} \approx 3.6 \times 10^{38}$ erg s$^{-1}$. This makes AT2019dsg relatively radio-quiet \cite[e.g.][]{2020SSRv..216...81A} and comparable to ASASSN-14li \citep{2016ApJ...819L..25A,2020SSRv..216...81A} and CNSS J0019+00 \citep{2020ApJ...903..116A}, whose radio emission is thought to be associated with non-relativistic outflows. 

For afterglow emission from a relativistically expanding region, the expected time-dependent angular size is $\theta_A \propto (E_{j,52}/n_0)^{1/4} t^{1/4}$ and $\propto (E_{j,52}/n_0)^{1/2} t^{1/2}$ for a constant density and a stratified surrounding medium respectively (see eqns. \ref{thetaAk0} and \ref{thetaAk2}), where $E_{j,52}$ is the total kinetic energy of the outflow scaled in units of $10^{52}$ erg and $n_0$ is the number density at a distance of 1 pc from the central engine. For the VLBI sessions 1 and 3, the constrained size $\theta_{A,{\rm obs}}$ is $ \leq 0.17$ mas and $\leq 0.44$ mas, respectively. On comparison with the above models, the best estimate gives $(E_{j,52}/n_0) \leq 0.08$ for session 1 (200 days post TDE) and a consequent Lorentz factor $\Gamma_{\rm sh} \leq 0.80$ (see eqns. \ref{gammashk0} and \ref{gammashk2}), indicative of a non-relativistic expansion in AT2019dsg, consistent with the above expectation.  
The VLBI measured flux densities and constraint on source size 
thus play a complementary role in enabling inferences on the emission source nature, and helping reduce the number of free parameters in the synchrotron radiation model. The lack of VLBI/high resolution radio observations during the early epochs, and the consequent unavailability of astrometric accuracy necessitates the use of a model based approach to infer the possibility of a relativistic expansion.  
 This may last for $\leqslant$ 18 d (relativistic phase marked by a Lorentz factor $\Gamma_{\rm} \leqslant 2$; see eqns. \ref{gammashk0} and \ref{gammashk2}) based on the above expansion models; a non-relativistic expansion is thus more likely at $\geqslant$ 18 d, and an appropriate radiative model can offer a better description of the compiled data points that sample later epochs ($\geqslant$ 55 d). The detection of a relativistic jet during the early expansion phase could be indirectly inferred through the modeling of the light curves \cite[e.g.][]{2018Sci...361..482M} . This is as the afterglow emission from the expanding shock resulting from the jet - CNM interaction can be substantiated by emission from internal shocks in the jet \cite[e.g.][]{2020SSRv..216...81A}. In the absence of clear signatures of relativistic beaming based on low brightness temperatures, the emission from an off-axis jet may be significantly suppressed at large inclination angles, implying a domination by the afterglow emission. If the jet can survive the passage through a dense CNM, or with a favorable line of sight that renders it off-axis with respect to the observer, it may be detected and resolved such as in the case of Arp 299B AT1 at late epochs \citep{2018Sci...361..482M}. The significantly larger distance of AT2019dsg (in relation to Arp 299B AT1), and the decreasing flux density during the VLBI epochs however pose impediments to the detection.

The total energy in the outflow is minimal when the energy density of the relativistic electrons accelerated by the forward shock and that of the magnetic field are in equipartition \citep{1977MNRAS.180..539S,1998ApJ...499..810C}. The peak flux density is assumed to be associated with the transition of the synchrotron self-absorption frequency $\nu_a$ through the observation band. The evaluated peak flux density and associated time are then used to estimate the minimal size, magnetic field strength, energy output and ambient medium number density; all of these parameters are critical for a non-relativistic expansion \citep{1998ApJ...499..810C,2020ApJ...903..116A}. The application involves a sensitive dependence on microphysical parameters including the fraction of the total energy density in the electron kinetic energy $\epsilon_e$ and magnetic field $\epsilon_B$. We use $(\epsilon_e,\epsilon_B)$ of $(1/3,1/3)$ and $(0.1,0.01)$ as typically representative of TDEs and AT2019dsg, respectively \cite[e.g.][]{2019ApJ...871...73H,2020ApJ...903..116A,2021arXiv210306299C}.

The frequencies that mark transitions in the evolving spectrum and their time evolution are (see Appendix \ref{aftergloweq}): the self-absorption frequency $\nu_a \propto t^{-2 \alpha (p+5)/(p+4)}$, the synchrotron frequency $\nu_m \propto t^{-2 \alpha}$, and the synchrotron cooling frequency $\nu_c  \propto t^{6 \alpha -2}$. The ordering $\nu_m \leq \nu_a$ as expected for a non-relativistic outflow requires that the microphysical parameters $\epsilon_e = 0.1$ and $\epsilon_B \leq 0.01$ at $t = t_p$. With these dependencies and using the above microphysical parameters, the frequencies at $t = t_p$ are derived as $\nu_a = 4.59 \times 10^{9}$ Hz, $\nu_m \leq 2.56 \times 10^{9}$ Hz and $\nu_c \geq 1.92 \times 10^{12}$ Hz. This suggests that a spectral regime change to $\nu_m < \nu_a$ occurred well before or near the peak of the light curve from an initial ordering $\nu_a < \nu_m < \nu_c$ with the observation frequency $\nu$ transitioning to the optically thin regime immediately after. The minimal values of the equipartition size is $R_{eq} = 4.7 \times 10^{16}$ cm ($\sim 3 \times 10^{4}~R_S$, where $R_S = 1.59 \times 10^{12}$~cm is the Schwarzschild radius of a $5.4 \times 10^6 M_\odot$ SMBH), magnetic field strength $B_{eq} = 0.17$ G, energy output $E_{eq} = 4.9 \times 10^{49}$ erg, and the ambient number density is $n_{eq} = 5.7 \times 10^3$ cm$^{-3}$ (see Appendix \ref{aftergloweq}), indicating an energetic TDE \cite[compared to thermal dominated TDEs, e.g.][]{2020SSRv..216...81A}.  
As the outflow is inferred to accelerate in the early phase \citep{2021NatAs...5..510S}, it can have a constant velocity in a short phase between the flux density peak ($t_p = 152.8$ d) and the first VLBI epoch (200 d), with a lower limit of $R_{eq}/t_p \approx 0.12~c$ for $t_p = 152.8$ d.

The partial disruption of a star (50 \%) with mass $M_\star$ can potentially produce an energy $E_\star = \eta (M_\star/2) c^2$. Assuming an accretion conversion efficiency $\eta = 0.01$ and equating with the energy output in the outflow (eqn. \ref{Ejeq} with a SMBH mass $M_\bullet = 5.4 \times 10^6$ $M_\odot$), we obtain a stellar mass in the range $0.14 - 2.05$ $M_\odot$ for a penetration factor $\beta$ in the range $1 - 7.86 M^{-2/3}_\bullet (M_\star/M_\odot)^{-0.07}$ (see Appendix \ref{fbacc}). The corresponding $\beta$ is $1 - 2.9$, outflow energy is $\leqslant 1.8 \times 10^{52}$ erg and initial outflow velocity is $0.25 - 0.39\, c$, consistent with the expected deceleration to 0.12 $c$. The initial mass outflow rate is $(3.1 - 4.9) \times 10^{-2}$ $M_\odot$ yr$^{-1}$ (for $t = t_p$). The slowdown is likely enabled by the outflow interaction with the surrounding CNM, owing to a density contrast ratio between the CNM and outflow of $\geq$ 32.3 at $t = t_p$ indicating a denser CNM (using eqn. \ref{kappaeqn}). This could be aided by contributions from radiation drag, especially for high accretion rates and in the vicinity of the SMBH, which acts to make the outflow radially directed and at a saturated constant velocity \cite[e.g.][]{2015ApJ...805...91M,2021MNRAS.500.2620Y}.  

The outflow can originate either from a wind driven by the radiation pressure of the accretion disk \cite[e.g.][]{2015ApJ...805...91M,2021MNRAS.500.2620Y} or from unbound debris from the stellar disruption \cite[e.g.][]{2016ApJ...827..127K}. In the former scenario, the lifetime is between 40.9 d and 228.0 d, set by the requirement that the shrinking photospheric radius equals the accretion disk size ($\approx$ 10 $R_S$, $R_S$ is the Schwarzschild radius of the black hole), using eqn. \ref{Rph}. The corresponding disk luminosity is $\sim 0.2$ times the Eddington luminosity which can support non-relativistic winds with velocities $\leqslant 0.2~c$ driven by radiation pressure from the disk emission \cite[e.g.][]{2021MNRAS.500.2620Y}. In the latter scenario, comparing the outflow velocity inferred here ($\approx 0.12~c$) to that associated with the outflow \citep{2016ApJ...827..127K} yields a non-physical stellar mass of $\geqslant 1.3 \times 10^{13} M_\star$. An origin as radiation-driven winds is thus more likely.

The acceleration of ultra-high energy cosmic rays \cite[e.g.][]{2009ApJ...693..329F} is closely associated with mechanisms of neutrino production \cite[e.g.][]{2018A&A...616A.179G}. This is naturally enabled in the case of a relativistic jet, through hadronic and photo-hadronic interactions \cite[e.g.][]{2018A&A...616A.179G,2018NatSR...810828B}. In the proposed concordance scenario \citep{2021NatAs...5..472W}, the neutrino production is attributed to the interaction between X-ray photons (back-scattered from an expanding optically thick cocoon) and protons accelerated by internal shocks in a relativistic jet. As the density contrast is not especially large, a dark or hidden relativistic jet appears unlikely, also owing to the non-detection of a significant proper motion. For a relativistic jet with a bulk Lorentz factor equal to the Doppler boosting factor, the apparently superluminal velocity $v_\perp$ is related to the jet viewing angle $\theta$ as $(v_\perp/c) = 1/\tan \theta$. For a $\theta = 10^\circ - 45^\circ$, $v_\perp = 1 - 5.67~c$ corresponds to a proper motion of $0.29 - 1.67$ mas yr$^{-1}$ which exceeds our measurement of $\leq$ 0.16 mas yr$^{-1}$ thus rendering the proposed scenario for neutrino production by photo-hadronic interactions in a relativistic jet less likely. 

We then explore non-jetted scenarios enabling the production of neutrinos. These include an association with the outflow or the accretion process \cite[e.g.][]{2019ApJ...886..114H,2020ApJ...902..108M}, both of which are potential candidates but neither can achieve the expected all-neutrino fluence \citep{2020ApJ...902..108M}. A detailed elaboration of the neutrino production is beyond the current scope. We thus discuss the relative plausibility of these scenarios broadly by estimating the maximum energy to which the protons (constituting the cosmic rays which produce the neutrinos through various reaction channels) can be accelerated and the associated timescales. In the former (outflow), the protons are accelerated by the outflow and the maximum energy \cite[e.g.][]{2017PhRvD..96f3007Z} is $E_p \propto t^{-1} \leqslant 1.3 \times 10^{16}$ eV (at a time $t =$ 175 d), limited by the dynamic timescale (see eqn. \ref{Epoutflow}). The detected neutrino energy $E_\nu \approx 1.5$ \% of $E_p$ indicating that this is a possible channel. In the latter (accretion), the super-Eddington regime (with a lifetime $t_j \approx 35.3 - 576.3$ d) which follows the debris circularization ($\leq 11.1$ d from eqn. \ref{tcirc}) is found to be inefficient in producing relativistic protons as the cooling from proton-proton collisions effectively limits the maximum energy \citep{2019ApJ...886..114H}. In the radiatively inefficient accretion regime, the acceleration of protons by plasma turbulence is limited by their diffusion with a consequent maximum energy $E_p \leq 3.7 \times 10^{16}$ eV (see eqn. \ref{EpRIAF}) with the neutrino energy being $\approx 0.5$ \% of $E_p$. However, the timescale for transition from the super-Eddington to radiatively inefficient accretion flow is long ($\approx 16 t_j \geq 5.65 \times 10^{2}$ d), rendering it unlikely to occur during the current observation period. We thus tentatively favour the outflow scenario for the production of neutrinos.

The neutrino production itself may be enabled by either the $p\gamma$ or $pp$ interactions at the base of the outflow, 
 taken to be the photospheric radius, $R_{ph}$. By equating $R_{ph}$ from eqn. \ref{Rph} with an accretion disk size  $\approx 10~R_S$ (photosphere outside but in the vicinity of the disk), the resultant solution of $R_{ph}/(10~R_S) \approx 1.56$ constrains the stellar mass $M_\star$  in the range $1.36 - 2.05$ $M_\odot$ with a penetration factor $\beta = 1$ indicating that the disruption involves an extremely close approach to the SMBH. The photosphere is just outside and likely to be spatially distributed \cite[e.g.][]{2009MNRAS.400.2070S} around the disk. This site provides a suitable cross section for the interactions with a particle number density $\geq 10^{8}$ cm$^{-3}$ \citep{2021NatAs...5..510S} for $\gamma \geq 1.3$ assuming that the CNM density scales as $R^{-\gamma}$ in the innermost region. Furthermore, at the same epoch, the number density of unbound stellar debris in the vicinity of the accretion disk is $\leq 3.9 \times 10^{11}$ cm$^{-3}$  using eqn. \ref{naeqn} thus providing support to this scenario. A production at the radio emission region at $R_{eq} \approx 5.1 \times 10^{16} $ cm is unlikely owing to a relatively low particle number density $n_{eq} = 4.1 \times 10^3$ cm$^{-3}$.

The VLBI observations cover the declining phase of the source evolution, and thus provide unique constraints on the flux density and proper motion. These play a key role in inferring a decelerating non-relativistic expansion, and in reducing the free parameters in the radiative model. 
Similar to CNSS J0019+00 and ASASSN-14li, a relatively radio-quiet ($\approx 4 \times 10^{38}$ erg s$^{-1}$) nature of AT2019dsg with a predominantly thermal emission and the non-detection of a relativistic jet are consistent with that the non-thermal radio emission is from outflow activity. The large density contrast ($\geq 32.3$), albeit during the later epochs, hints at an outflow that may have decelerated owing to a 
frustration by the CNM \citep{2021NatAs...5..510S}.

\section{Summary}

Using the results from the present study together with the measured flux densities during the rising phase in literature, the following scenario emerges: a star of $\approx 1.35 - 2.05~M_\odot$ is disrupted; a non-relativistic outflow with a velocity of $\approx 0.1~c$, originating from and radiatively driven by gas accretion onto the supermassive black hole, interacts with the surrounding CNM at $\geq 4.7 \times 10^{16}$ cm ($\sim 3 \times 10^4~R_S$); the interactions result in an expanding forward shock that accelerates ambient electrons (constituting the CNM) to relativistic energies, thus producing the observed synchrotron radio emission; the production of neutrinos is likely at the base of the outflow, where protons can be efficiently accelerated to Peta eV energies and provide suitable cross sections for hadronic or photo-hadronic interactions, which is consistent with such an expectation from a detailed analysis in \citet{2021NatAs...5..510S}. The origin of neutrinos inferred here is then in conflict with a possible origin from internal shocks in a relativistic jet \citep{2021NatAs...5..472W} or even a non-expectation as espoused by \cite{2021arXiv210306299C}.

As the VLBI observations offer high resolutions, they 
offer a complementary (compared to multi-wavelength electro-magnetic observations) but unique perspective on the source evolution. Studies of TDEs, especially those nearby, are crucial to probe the outflow nature and constrain the CNM properties of the galactic nuclear regions. It is crucial that the VLBI monitoring of TDEs in general be initiated during the early evolutionary phases of the source evolution. This will help capture any possible relativistic expansion and the subsequent transition to a non-relativistic phase. A sparse but active monitoring during later epochs can help identify the presence of a possible jet, if the conditions (nearby source, activity of the central engine, density of the surrounding medium) are favourable for such an inference. This information is highly relevant for the accurate modelling of the afterglow radiative evolution that enables the extraction of physical parameters pertaining to the TDE physical and geometrical properties.  
The inference of a non-relativistic outflow, and its potential role in the production of neutrinos provides an expanded inventory of extra-galactic sources of neutrinos other than the Galactic PeV accelerators \citep{2016Natur.531..476H,2021Innov...200118G}. This includes blazar jets \cite[e.g.][]{1993A&A...269...67M,2020MNRAS.497..865G}, the dominant non-blazar AGN populations \cite[e.g.][]{2016PhRvD..94j3006M,2019JCAP...02..012H} and slow transients including gamma-ray bursts \cite[e.g.][]{1998PhRvD..58l3005R, 1999PhRvD..59b3002W}, making multi-messenger investigations of similar sources essential and exciting \cite[e.g.][]{2017A&A...603A..76G}. This could be relevant in planning strategies to enable the effective use of global facilities for multi-messenger astronomy. 


\begin{acknowledgments}
We thank the anonymous referee for the constructive comments and suggestions, the inclusion of which have improved the content and presentation of our work.
This work is supported by National Key R\&D Programme of China (grant number 2018YFA0404603) and NSFC (1204130).  
 The European VLBI Network (EVN) is a joint facility of independent European, African, Asian, and North American radio astronomy institutes. Scientific results from data presented in this publication are derived from the following EVN project code(s): EM140, RSM04. We thank Wen Chen at Kunming Station for making the first observation.
 e-VLBI research infrastructure in Europe is supported by the European Union, Seventh Framework Programme (FP7/2007-2013) under grant agreement number RI-261525 NEXPReS.
 e-MERLIN is a National Facility operated by the University of Manchester at Jodrell Bank Observatory on behalf of STFC. The VLBI data processing made use of the compute resource of the China SKA Regional center prototype, funded by the Ministry of Science and Technology of China and the Chinese Academy of Sciences. 
\end{acknowledgments}

\begin{appendix}

\section{VLBI observations and data reduction.}\label{VLBIobs}

The 5-GHz European VLBI network (EVN) observations were carried out on 2019 October 27 and November 12, and 2020 February 28.
The observational setup is presented in Table \ref{tab:obs}. It includes the EVN project code, date, frequency, time duration, bandwidth and participating telescopes (which included a maximum of 20 VLBI telescopes). The longest baselines in the East-West (Yebes, Spain — Tianma, China) and North-South (Badary, Russia - Hartebeesthoek, South Africa) directions span up to $\approx$ 9000 km and 9800 km, respectively, contributing to the mas-scale high resolution. The large number of participating stations provided a good sampling in the ($u, v$) plane, optimally covering the source and a resulting high-sensitivity image with a root-mean-square (rms) noise down to $\approx$ 0.015~mJy~beam$^{-1}$. 
 
All the observations were conducted by utilising the phase-referencing technique \citep{1995ASPC...82..327B}. During the observations, J2052$+$1619 served as the primary phase-referencing calibrator (\texttt{Cal1}, $\sim$200 mJy at 5 GHz), 2.37 degree away from the target source AT2019dsg (\texttt{Tar}) on the plane of the sky. The nearby 20-mJy source J2058$+$1512 was used as the secondary phase calibrator (1.08 degree away from the target). The bright ($\sim$1 Jy) radio source J2031$+$1219 was observed as the fringe finder. The estimated VLBI flux densities of the above three calibrators are all obtained from the ASTROGEO Center\footnote{$http://www.astrogeo.org$}, where accurate astrometry information can also be found for selecting appropriate calibrators. The main observing cycle is “\texttt{Cal1} (60s) — \texttt{Tar} (200s) — \texttt{Cal1} (60s)”.
The secondary phase-referencing calibrator was observed for a scan of 200 s every five main cycles, serving as a reference to check the accuracy of the phase-referencing results. In sessions 1 and 3, the observational data at each station were recorded independently in disk modules and shipped to the correlator center hosted by the Joint Institute for VLBI ERIC (JIVE), located at Dwingeloo, the Netherlands. The total data recording rate is 2 gigabits per second. The disk packs from all stations were shipped to the EVN correlator SFXC \citep{2015ExA....39..259K}. The correlation was carried out using the typical correlation setups for continuum sources (\textit{i.e.}, 1 s integration time and 1 MHz frequency resolution). In the second session (code: RSM04), the data recorded at each station were directly transferred in real time via broad-band optical fibre to the SFXC software correlator in JIVE. Correlation was also carried out in real time mode. The rapid correlation and release of the second-session data enable us to quickly obtain the radio image of AT2019dsg 216 days post burst, validating the successful detection from the VLBI observation. 
 
The correlated data were calibrated by following the standard procedures using the NRAO Astronomical Image Processing System (AIPS) software \citep{2003ASSL..285..109G}. After loading the data with task FITLD, the interferometric visibility amplitudes were calibrated using the system temperatures and antenna gain curves measured at each telescope using task ANTAB. For some antennas (EM140A: Km; RSM04: Bd, Ir, Da, Kn, Pi) without system temperature records during the observations, we initially used the nominal values of the system equivalent flux density presented in the EVN status table\footnote{\url{http://old.evlbi.org/user_guide/EVNstatus.txt}} instead. The beginning 30 s of each scan were flagged due to the occupation of antenna slewing. The ionospheric dispersive delays were corrected using a map of total electron content provided by Global Positioning System satellite observations\footnote{the model files were downloaded from https://cddis.nasa.gov/archive/gps/products/ionex} and applied to the visibilities through task TECOR. The VLBA procedure task VLBAPANG was used to correct phase variations caused by parallactic angles. A manual phase calibration and bandpass calibration were carried out using the best scans of the fringe finder through task FRING and BPASS. The global fringe-fitting was then performed on the phase-referencing calibrator J2052$+$1619.

After fringe-fitting of the primary phase-referencing calibrator, their visibility data were exported from AIPS and imported into Caltech Difmap software package \citep{1997ASPC..125...77S} for self-calibration and imaging. For these antennas, direct use of the nominal T$_{\rm sys}$ values could result in a large uncertainty on the source amplitude calibration. To avoid this issue, we first built CLEAN models of the phase-referencing calibrator only using antennas with T$_{\rm sys}$ measurements. After good gain solutions were obtained, we fixed the amplitude gain factors of these antenna by setting them to the reference antenna; next, we ran amplitude and phase self-calibration again for all antennas (\textit{i.e.}, including those without T$_{\rm sys}$ values). Solving the closure amplitude then yielded the amplitude scaling factors for antennas with nominal T$_{\rm sys}$ values. The corresponding antenna gain corrections from amplitude self-calibration were then applied to the data using AIPS task CLCOR. Additional iteration of fringe-fitting was performed on the phase-referencing calibrator by taking account of its CLEAN component model to remove the small phase variations due to the core-jet structure of the phase calibrator. Then the obtained solutions were finally interpolated and applied to the secondary calibrator and the target source using CLCAL. The calibrated data were split into single-source files and imported into Difmap for imaging and model-fitting.
The above data reduction and imaging procedures are the same for all sessions. 
During the data processing, we found that some antennas have relatively large phase errors due to the coherent loss or bad weather, we thus excluded these antennas (EM140A: Sv, Zc, Hh, Wb; RSM04: Zc, Hh, Wb, Cm, De; EM140B: Sv, Zc, Hh, Wb). 
The participating antennas are mentioned in Table \ref{tab:obs}.
Given the weakness of the target sources, no self-calibration was conducted on it. After CLEANing the peak regions of the map (\textit{i.e.}, signal-to-noise level $\ge$ 5$\sigma$), natural grid weighting was used to create the final images. 
We used the Modelfit procedure in Difmap to constrain the observing parameters of the source (\textit{e.g.}, positions, sizes, flux densities). Circular Gaussian models were used in the model fitting. Table \ref{tab:parm} presents the fitting results. Uncertainties in the parameters are calculated based on that expected from an imaging analysis \citep{1999ASPC..180..301F} and include additional inputs.
For the peak intensity and integrated flux density, an extra uncertainty of 5 per cent was coupled with the original errors. The size errors are derived by using the ratio of the minor FWHM (full width at half maximum) size of the synthesised beam to the component SNR (peak intensities divide by r.m.s noises). The following equations are used to estimate the above uncertainties \cite[e.g.][]{1999ASPC..180..301F}
\begin{align}
\sigma_{\rm peak}&=\sqrt{\sigma_{\rm rms}^2 + (0.05~S_{\rm peak})^2} \\
\sigma_{\rm tot}&=\sqrt{\sigma_{\rm rms}^2+\left(\frac{S_{\rm tot}}{\rm SNR}\right)^2+(0.05~S_{\rm tot})^2} \\
\sigma_{\theta}&=\frac{\theta_{\rm FWHM}}{\rm SNR}
\end{align}
Where $\sigma_{\rm rms}$ is the r.m.s noise of the images and SNR is the signal-to-noise ratio $\left(\dfrac{S_{\rm peak}}{\sigma_{\rm rms}}\right)$, and $\sigma_{\rm peak}$, $\sigma_{\rm tot}$ and $\sigma_{\theta}$ are the uncertainties of the peak flux density, the integrated flux density and the FWHM size of a fitted Gaussian component, respectively.

The first full-array 8-hr EVN observation was conducted on 2019 October 27 (200 days after TDE). It was proposed as part of a two-epoch monitoring program of AT2019dsg to monitor the flux density and size evolution of the afterglow and to infer possible proper motion. A compact source with an integrated flux density of 0.61 mJy was constrained to a size $\leq$ 0.17 mas. 

Following this, a target of opportunity (ToO) 3-hr \textit{e}-EVN observation was conducted on 2019 November 12 (216 days after the TDE). The aim was to verify the compactness of the source 
and the potential for long-term monitoring of the source. 
A compact source was successfully detected from this fast-response observation with an integrated flux density of 0.674$\pm$0.048 mJy. 
The peak flux density is $\approx$ 28 times the \textit{rms} noise, providing reliable detection confidence in the results from the first EVN session and detection promise for the next full-disk observation session scheduled three months later. 

The second full-array 8-hr EVN observation was carried out on 2020 February 28 (session 3; 304 days after TDE),  mainly to infer any possible proper motion, if the source remained detectable. A compact source was again detected, although its size expanded to $\leq$ 0.48 mas and its integrated flux density was reduced to 0.44 mJy. 

The astrometric position derived by averaging the peak positions over the three epochs is $\alpha$(J2000) = 20h 57m 02.96434s $\pm$ 0.00001s and $\delta$(J2000) = 14$^{\circ}$12$^\prime$16.29425$^{\prime\prime}$ $\pm$ 0.00020$^{\prime\prime}$. The uncertainties are the square root of the sum of the squares of two error terms: the astrometric error of phase-referencing calibrator Cal1 (0.13 mas on right ascension and 0.17 mas on declination, from ASTROGEO) and the simulated astrometric accuracy of VLBI phase-referenced observations (that gives a statistical positional error of $\Delta_{RA}$ $\approx$ 0.06 mas and $\Delta_{DEC}$ $\approx$ 0.10 mas; see \citealt{2006A&A...452.1099P}). The total uncertainty is dominated by the astrometric accuracy of the calibrator.

Errors on the measured source positions are composed of systematic and random contributions. 
Systematic errors include contributions from instrumental and imaging sources. We followed a similar approach to that of \cite{2020ApJ...888L..24M}, with the employment of two phase calibrators CAL1 (J2052+1619) and CAL2 (J2058+1512). 
To minimize systematic errors, in each epoch, we used the calibration solutions from CAL1 to derive the positions for both CAL2 and the target source (AT2019dsg). Then we calculated the relative positions between AT2019dsg and CAL2 (see Column 5 in Table \ref{tab:err}); this helped monitor positional changes 
across the three observation epochs. 
Since CAL2 is an extra-galactic radio quasar, and the positions used correspond the the bright radio core, 
CAL2 can be treated as a stable source across the three epochs. The above methodology renders most instrumental and calibration based errors negligible. The remaining contributors to systematic errors are then only from the measurement errors on CAL2 and AT2019dsg, as well as the phase referencing errors (i.e. the error caused by the separation between AT2019dsg and the phase calibrator, see \citealt{2006A&A...452.1099P} for details). The measurements and errors are listed in Table \ref{tab:err}. 
As we use the relative position (w.r.t. CAL2) to identify any shift in the emission center of AT2019dsg, a major portion of the systematic errors can be eliminated. 
The systematic errors $\sigma_{sys}$ reported in Column (6) of Table \ref{tab:err} are estimated as the standard deviation on the CAL2 offsets during the three observation epochs. For the random error $\sigma_{ran}$, a fitting error similar to that suggested in \cite{1999ASPC..180..301F} was applied, with $\sigma_{ran} = \theta_{\rm FWHM}/(2~{\rm SNR})$. For resolved components, the $\theta_{\rm FWHM}$ are the fitted component sizes; for unresolved sources, $\theta_{\rm FWHM}$ are the synthesised beams. In columns 7, the random errors for both the target and CAL2 were involved. The final errors (columns 8) were calculated by combining the error from both contributions (columns 6 and 7). Using these measurements (Columns 5 and 8 of Table \ref{tab:err}), the relative position change of AT2019dsg w.r.t. CAL2 is found to lie at a mid-point $({\rm RA},{\rm DEC}) = (0.06 \pm 0.12~{\rm mas}, -0.07 \pm 0.13~{\rm mas})$ ($1-\sigma$ uncertainties; see Fig. \ref{tar-img} panel 4 which illustrates these measurements and the $3-\sigma$ uncertainty ellipse enclosing them) indicating that there is no significant shift in the emission center of AT2019dsg across the three observational epochs. The distance between the mid-point and any of the other two points sets a maximum relative position change of $0.32 \pm 0.22$ mas ($1-\sigma$).  
\begin{table}
    \small
    \caption{EVN 5-GHz observations of AT2019dsg.}
    \begin{tabular}{cccccccc} \hline \hline 
Session & Date & MJD & Array & $\tau_\mathrm{int}$ & $\theta_{\rm beam}$ & $\sigma$ & Telescopes \\ 
     & &  (day)    &       & (min)       & (mas, mas, $^\circ$) & (mJy beam$^{-1}$) &    \\  
\hline 
EM140A & 2019 10 27 & 58783.6 &full EVN$^{1}$ &255 &6.1$\times$0.9 at 9.6$^{\circ}$ & 0.015&  Out: Ir\\ 
& & & & & & &Flag: Sv,Zc,Hh,Wb\\
RSM04  & 2019 11 12 & 58799.6 &\textit{e}-EVN$^{2}$ &105 &8.6$\times$2.3 at $-$13.2$^{\circ}$ &0.024 & Out: Sv, Ur, Km \\
& & & & & & &Flag: Zc,Hh,Wb,Cm,De\\
EM140B & 2020 02 28 & 58907.3 &full EVN &255 &5.3$\times$1.3 at 6.5$^{\circ}$ &0.016 &Out: Ur, Km  \\
& & & & & & &Flag: Sv,Zc,Hh,Wb\\
\hline 
    \end{tabular} \\[2mm]
Note: Column 1 – Observation session code; Column 2 – Observation date; Column 3–Observation time in Modified Julian Dates (MJD); Column 4 – VLBI network; Column 5 – Integration time of the target; Column 6 – Synthesised beam size (major, minor and position angle of the major axis); Column 7 – Post-clean \textit{rms} noise in the image; Column 8 – Notifications for participating antennas (Out: The antennas did not join the observation).
\\
$^{1}$EVN telescopes participating the observation are: Bd (Badary 32m, Russia), Ef (Effelsberg 100m, Germany), Hh (Hartebeesthoek 26m, South Africa), Ir (Irbene 16m, Latvia), Jb (Lovell 76m, UK), Km (Kunming 40m, China),  Mc (Medicina  25m, Italy), On85 (Onsala 25m, Sweden), Sv (Svetloe 32m, Russia), T6 (Tianma 65m, China), Tr (Torun 32m, Poland), Ur (Urumqi 26m, China), Wb (Westerbork 25m, Netherlands), Ys (Yebes 40m, Spain), Zc (Zelenchukskaya 32m, Russia). 
\\
$^{2}$For \textit{e}-EVN, 5 extra antennas from \textit{e}-MERLIN are included in the above array: Cm (Cambridge 32m, UK), Da (Darnhall 25m, UK), De (Defford 25m, UK), Kn (Knockin 25m, UK), Pi (Pickmere 25m, UK).
\label{tab:obs} 
\end{table}

\begin{table*}
\centering
    \caption{Target and calibrator astrometric information (Unit: mas)
    }
\resizebox{\columnwidth}{!}{
    \begin{tabular}{c|cc|c|cc|cc|cc|cc|cc} \hline \hline 
Session & \multicolumn{2}{|c|}{CAL2 offsets} & Target peak pos. & \multicolumn{2}{c}{Target offsets} & \multicolumn{2}{|c|}{Relative positional change} &\multicolumn{2}{c}{$\sigma_{sys}$} &\multicolumn{2}{|c}{$\sigma_{ran}$}& \multicolumn{2}{|c}{$\sigma_{tot}$}  \\
  & RA & DEC & (J2000) & RA & DEC & RA & DEC &RA & DEC &RA & DEC&RA & DEC  \\ \hline
 (1)&\multicolumn{2}{c|}{(2)}&(3) &\multicolumn{2}{c|}{(4)}&\multicolumn{2}{c|}{(5)}&\multicolumn{2}{c|}{(6)}&\multicolumn{2}{c|}{(7)}&\multicolumn{2}{c}{(8)} \\
\hline
EM140A & 3.99 &3.09 & 20:57:02.96434$+$14:12:16.29418 &$-$9.88&4.34&0.00&0.00&0.16&0.13&0.02&0.08&0.16&0.15 \\
RSM04 & 3.67 &3.41 & 20:57:02.96434$+$14:12:16.29414 &$-$9.89&4.39&0.31&-0.27&0.16&0.13&0.06&0.16&0.17&0.20 \\
EM140B & 4.04 &3.28 & 20:57:02.96433$+$14:12:16.29444 &$-$10.02&4.66&-0.19&0.13&0.16&0.13&0.03&0.10&0.16&0.16 \\

\hline 
    \end{tabular}%
}
    \\[2mm]
Note: Column 1– session code; Columns 2– Offsets of the measured CAL2 positions with respect to the phase center of CAL2 (20:58:45.819$+$15:12:13.186); Column 3- target peak position;  
Columns 4– The offsets of the measured target positions with respect to the phase center of target source (20:57:02.965$+$14:12:16.290); 
Columns 5- The relative position between the target source and CAL2. To constrain the positional change of the target source, the first epoch was set to (0,0), and the other two epochs are in reference to the first epoch;
Columns 6– Systematic errors on the target source relative positional change, estimated from the standard deviations of the CAL2 offsets (Column 2);
Columns 7- Random error on the target source relative positional change, estimated from the fitting errors \cite[e.g.][]{1999ASPC..180..301F}; 
Columns 8- Total error on the target source relative positional change based on Columns 6 and 7.
\label{tab:err}
\end{table*}

\begin{table*}
    \centering
    \small
    \caption{Imaging parameters of AT2019dsg}
    \begin{tabular}{ccccccc} \hline \hline 
Session & MJD & t$_{\rm post}$ & S$_{\rm peak}$ & S$_{\rm int}$ & $\theta_{\rm FWHM}$ & T$_{\rm B}$ \\
     &  (day)    & (day)      & (mJy beam$^{-1}$)       & (mJy) & (mas)  &(K)    \\  
\hline 
EM140A & 58783.6 & 200.0 & 0.598$\pm$0.034  & 0.608$\pm$0.037 & 0.17$\pm$0.02 & 1.1$\times10^{9}$   \\
RSM04 & 58799.6 & 216.0 & 0.651$\pm$0.040 & 0.674$\pm$0.048 & $\le$0.60 & $\ge$9.8$\times10^{7}$ \\
EM140B  & 58907.3 & 323.7 & 0.442$\pm$0.027 & 0.442$\pm$0.032 & $\le$0.48 &     $\ge$1.0$\times$10$^{8}$ \\

\hline 
    \end{tabular} \\[2mm]
Note: Column 1–Session code; Column 2–Observation time in Modified Julian Dates (MJD); Column 3–Day after the TDE burst; Column 4–Peak intensity of the detected source; Column 5–Integrated flux density of the fitted Gaussian component; Column 6–The full width at half maximum (FWHM) size of the fitted circular Gaussian component; Column 7–Brightness temperature of the fitted component.
\label{tab:parm} 
\end{table*}

\section{Light curve fitting}\label{lcfitting}

The compiled 5-GHz light curve (see Figure \ref{lcurvefit}) includes data points covering the rising phase and peak \citep{2021NatAs...5..510S,2021MNRAS.504..792C}, and our three VLBI points in the declining phase. Data points from \citep{2021arXiv210306299C} are not employed in the fit owing to their VLA flux densities being systematically lower than the e-MERLIN measurements reported in \cite{2021MNRAS.504..792C}. 
As our VLBI estimates are likely to be closer to the e-MERLIN values, we mainly use the former measurements only when conducting the light curve fitting. 

The light curve is fitted using the weighted non-linear least squares method with a model of the form $F_\nu = F_{\nu_0}+F_{\nu_1} ((t/t_p)^{5 \alpha_1}+(t/t_p)^{5 \alpha_2})^{-1/5}$, with weights based on the flux density measurement errors, the flux density normalization $0.0 < F_{\nu_0} \leq 0.5$ mJy, $0.0 < F_{\nu_1} \leq 3.0$ mJy, the rising slope $0.0 \leq \alpha_1 \leq 5.0$ and declining slope $-10.0 \leq \alpha_2 \leq 0.0$, a turnover time $t_p$ days and a smoothness parameter fixed at 5. 
 The availability of flux density measurements during the last three epochs based on our EVN observations plays a crucial role in constraining the optically thin declining phase. The fit yields the parameters $F_{\nu_0} = 0.17 \pm 0.17$ mJy, $F_{\nu_1} = 1.18 \pm 0.18$ mJy, $\alpha_1 = 2.38 \pm 0.93$, $\alpha_2 = -2.56 \pm 1.31$. These yield a peak flux density $F_{\nu_p} = 1.19 \pm 0.18$ mJy and an associated peak time $t_p = 152.80 \pm 16.19$ days. The weighted nature of the fitting through the dependence on measurement errors in the data points is reflected in the relatively large uncertainties above. This is especially the case near the light curve peak (flux density and time of occurrence) and in the declining slope where a paucity of data points and measurement errors render larger imprints. The fit parameters are summarized in Table \ref{tab:fitpars}.
 \begin{table*}
   \centering
    \small
    \begin{tabular}{rrr} \hline \hline 
 Parameter & Range & Fit value \\  \hline 
Normalization (constant) $F_{\nu_0}$ (mJy) & 0.00 -- 0.50 & 0.17 $\pm$ 0.17\\
Normalization (shape) $F_{\nu_1}$ (mJy) & 0.00 -- 3.00 & 1.18 $\pm$ 0.18\\
Rising slope $\alpha_1$ & 0.00 -- 5.00 & 2.38 $\pm$ 0.93\\
Declining slope $\alpha_2$ & $-$10.00 -- 0.00 & $-$2.56 $\pm$ 1.31\\
Peak time (days) & 100.00 -- 300.00 & 152.80 $\pm$ 16.19\\
Peak flux density $F_{\nu_p}$ (mJy) & - & 1.19 $\pm$ 0.18\\ \hline 
\end{tabular} \\[2mm]
\caption{Fit parameters for the model $F_\nu = F_{\nu_0}+F_{\nu_1} ((t/t_p)^{5 \alpha_1}+(t/t_p)^{5 \alpha_2})^{-1/5}$. The large uncertainties in the peak time and declining slope are a consequence of sparse data points in these evolution phases.}
\label{tab:fitpars} 
\end{table*}

 \section{Afterglow: hydrodynamics}\label{afterglowhydro}

The interaction of the outflow with the CNM produces an expanding forward shock. Assuming an initial relativistic expansion, the Lorentz factor of the shock $\Gamma_{\rm sh}$ and the projected size $r$ of the emitting region are often assumed to follow a self-similar evolution \citep{1976PhFl...19.1130B}, and are given by \cite[e.g.][]{2012MNRAS.420.3528M}
\begin{align}
\Gamma_{\rm sh} &= \left(\frac{17-4 k}{16 \pi} \frac{E_j}{m_p n r^3 c^2}\right)^{1/2} \label{gammash}\\ 
r &= \frac{2 \Gamma^2_{\rm sh} c t}{(1+z)}\label{rsh},
\end{align}
where the number density is assumed to scale with distance as $n = n_0 (r/r_0)^{-k}$ for a scaling constant $k = 0$ in the case of a surrounding medium of constant number density and $k = 2$ for a stratified medium, for a fiducial distance $r_0$ with associated number density $n_0$. The total kinetic energy of the outflow $E_j$. Using the scalings $E_{j,52} = E_j/(10^{52}~{\rm erg~s^{-1}})$, $t_d = t/(86400~{\rm s})$, $D_{L,100} = D_L/(100~{\rm Mpc})$ and assuming $r_0 = 1$ pc, the parameters $\Gamma_{\rm sh}$ and $\theta_A = r/D_A$ (where $\theta_A$ is the source angular size and $D_A = D_L/(1+z)^2$ is the source angular distance) are 
\begin{align}
\Gamma_{\rm sh} (k = 0) &= 7.96~\left(\frac{E_{j,52}}{n_0}\right)^{1/8} t^{-3/8}_d (1+z)^{3/8}\label{gammashk0}\\
\theta_A (k = 0) &= (0.18~{\rm mas})~\left(\frac{E_{j,52}}{n_0}\right)^{1/4} t^{1/4}_d (1+z)^{7/4} D^{-1}_{L,100}\label{thetaAk0}\\
\Gamma_{\rm sh} (k = 2) &= 2.21~\left(\frac{E_{j,52}}{n_0}\right)^{1/4} t^{-1/4}_d (1+z)^{1/4}\label{gammashk2}\\
\theta_A (k = 2) &= (0.02~{\rm mas})~\left(\frac{E_{j,52}}{n_0}\right)^{1/2} t^{1/2}_d (1+z)^{3/2} D^{-1}_{L,100}\label{thetaAk2},
\end{align}

\section{Afterglow: equipartition estimates and radiative evolution}\label{aftergloweq}

We assume that the radio afterglow emission originates from a non-relativistic spherically expanding forward shock. In this formulation \citep{1998ApJ...499..810C,2020ApJ...903..116A}, the peak of the light curve corresponds to the transition of the synchrotron emission from optically thick to thin through self-absorption. The relativistic electrons responsible for the radio emission are assumed to have a power law distribution in energy $E$, with the number per unit energy $N = N_0 E^{-p}$ where $N_0$ is a constant  evaluated at a given time and $p \geq 2$ is the power law index.

The flux densities in the optically thick ($\propto \nu^{5/2}$) and thin ($\propto \nu^{-(p-1)/2}$) regimes are
\begin{align}
F_\nu &= \frac{\pi R^2}{D^2_L} \frac{c_5}{c_6} B^{-1/2} \left(\frac{\nu}{2 c_1}\right)^{5/2} \label{fdopthick}\\
F_\nu &= \frac{4 \pi R^3 f}{3 D^2_L} c_5 N_0 B^{(p+1)/2} \left(\frac{\nu}{2 c_1}\right)^{-(p-1)/2} \label{fdopthin},
\end{align}
where $R$ is the size of the emitting region, $D_L$ is the luminosity distance, $B$ is the magnetic field strength, $f$ is the emission filling factor. The radiative constants $c_1$, $c_5$ and $c_6$ as functions of $p$ are given by \citep{1970ranp.book.....P}
\begin{align}
c_1 &= \frac{3 e}{4 \pi m^3_e c^5}\\
c_5 &= \frac{3^{1/2} e^3}{4 \pi m_e c^2 (p+1)} \Gamma\left(\frac{p}{4}+\frac{19}{12}\right) \Gamma\left(\frac{p}{4}-\frac{1}{12}\right)\\
c_6 &= \frac{3^{1/2} e^3}{8 \pi m_e} \left(\frac{3 e}{2 \pi m^3_e c^5}\right)^{-2} \Gamma\left(\frac{p}{4}+\frac{1}{6}\right) \Gamma\left(\frac{p}{4}+\frac{11}{6}\right)
\end{align}
where the electric charge $e = 4.8 \times 10^{-10}$ e.s.u., electron mass $m_e = 9.11 \times 10^{-28}$ g, speed of light $c = 3 \times 10^{10}$ cm s$^{-1}$, and $\Gamma$ represents the Gamma function. The constant energy density term $N_0$ in eqn. \ref{fdopthin} is obtained from the equipartition condition
\begin{equation}
N_0 = \left(\frac{\epsilon_e}{\epsilon_B}\right) \frac{B^2}{8 \pi} (p-2) E^{p-2}_l,
\label{N0}
\end{equation}
where $\epsilon_e$ and $\epsilon_B$ are the fractions of the total energy density in the particle kinetic energy density and in the magnetic field respectively, and $E_l = 0.51$ MeV $= 8.2 \times 10^{-7}$ erg is the electron rest mass energy and is representative of the energy at which the electrons transition to relativistic. 

Equating the flux densities from eqns. \ref{fdopthick} and \ref{fdopthin}, and using eqn. \ref{N0}, the minimal size  $R_{eq}$ and magnetic field strength $B_{eq}$ are expressed in terms of the peak flux density $F_{\nu_p}$ and corresponding frequency $\nu_p$ as 
\begin{align}
R_{eq} &= \left(\frac{6 c^{p+5}_6 F^{p+6}_{\nu_p} D^{2 p+12}_L}{(\epsilon_e/\epsilon_B) f (p-2) \pi^{p+5} c^{p+6}_5 E^{p-2}_l}\right)^{1/(2 p+13)} \left(\frac{\nu_p}{2 c_1}\right)^{-1}\\  \nonumber
&= (1.13 \times 10^{17}~{\rm cm}) \left(\frac{F_{\nu_p}}{\rm mJy}\right)^{0.47} \left(\frac{D_L}{100~{\rm Mpc}}\right)^{0.95} \left(\frac{\epsilon_e}{\epsilon_B}\right)^{-0.05} f^{-0.05} \left(\frac{\nu_p}{\rm GHz}\right)^{-1} \label{Req}\\
B_{eq} &= \left(\frac{36 \pi^3 c_5}{(\epsilon_e/\epsilon_B)^2 f^2 (p-2)^2 c^{3}_5 E^{2(p-2)}_l F_{\nu_p} D^2_L}\right)^{2/(2 p+13)} \left(\frac{\nu_p}{2 c_1}\right) \\ \nonumber
&= (0.06~{\rm G}) \left(\frac{F_{\nu_p}}{\rm mJy}\right)^{-0.11} \left(\frac{D_L}{100~{\rm Mpc}}\right)^{-0.22} \left(\frac{\epsilon_e}{\epsilon_B}\right)^{-0.22} f^{-0.22} \left(\frac{\nu_p}{\rm GHz}\right) \label{Beq},
\end{align}
and the minimal energy and number density are evaluated as
\begin{align}
E_{eq} &=\frac{B^2_{eq}}{8 \pi \epsilon_B} \frac{4 \pi}{3} R^3_{eq} f \\ \nonumber
&= (1.11 \times 10^{48}~{\rm erg}) \left(\frac{F_{\nu_p}}{\rm mJy}\right)^{1.20} \left(\frac{D_L}{100~{\rm Mpc}}\right)^{2.41} \left(\frac{\epsilon_e}{\epsilon_B}\right)^{-0.59} \epsilon^{-1}_B f^{0.41} \left(\frac{\nu_p}{\rm GHz}\right)^{-1} \label{Eeq}\\
 n_{eq} &= \frac{\epsilon_e}{\epsilon_B} \frac{B^2_{eq}}{8 \pi} \left(\frac{p-2}{p-1}\right) E^{-1}_l \\ \nonumber
&=(91.76~{\rm cm^{-3}}) \left(\frac{F_{\nu_p}}{\rm mJy}\right)^{-0.22} \left(\frac{D_L}{100~{\rm Mpc}}\right)^{-0.44} \left(\frac{\epsilon_e}{\epsilon_B}\right)^{0.56} f^{-0.44} \left(\frac{\nu_p}{\rm GHz}\right)^2  \label{neq},
\end{align}
where the expressions for $R_{eq}$, $B_{eq}$, $E_{eq}$ and $n_{eq}$ are evaluated for $p = 2.7$ and assuming fiducial values for the free parameters. 

The analysis assumes that the peak frequency $\nu_p$ is equal to the synchrotron self-absorption frequency $\nu_a$ which is obtained from the condition that the optical depth to self-absorption $\tau \approx \alpha_\nu R =$ 1 at $\nu = \nu_a$, where $\alpha_\nu$ is the absorption coefficient and $R$ is the time evolving size of the emission region. The time dependent physical parameters include $R \propto t^\alpha = R_{eq} (t/t_p)^\alpha$, $B \propto R^{-2} \propto t^{-2 \alpha} = B_{eq} (t/t_p)^{-2 \alpha}$ and $N_0 \propto B^2 = N_{0,eq} (t/t_p)^{-4 \alpha}$ where $N_{0,eq} = N_0 (B = B_{eq})$ (see eqn. \ref{N0}); $t_p$ is the time associated with the light curve peak.

The absorption coefficient is
\begin{equation}
\alpha_\nu = c_6 N_0 B^{(p+2)/2} \left(\frac{\nu}{2 c_1}\right)^{-(p+4)/2}.
\end{equation}
Using the condition $\alpha_\nu R = 1$ and the time dependent parameters, the synchrotron self-absorption frequency is 
\begin{align}
\nu_a &= 2 c_1 (R c_6 N_0)^{2/(p+4)} B^{(p+2)/(p+4)} \propto t^{-2 \alpha (p+5)/(p+4)}\\ \nonumber
&= (0.92 \times 10^{9}~{\rm Hz}) f^{-0.30} \left(\frac{\nu_p}{\rm GHz}\right) \left(\frac{t}{t_p}\right)^{-2.30 \alpha}.
\end{align}
The characteristic synchrotron frequency emitted by a relativistic electron with a Lorentz factor $\gamma_e$ is
\begin{equation}
\nu_e = \frac{e B \gamma^2_e}{2 \pi m_e c}.
\end{equation}
For $\gamma_e = \gamma_m = \displaystyle{\left(\frac{p-2}{p-1}\right) \left(\frac{m_p}{m_e}\right)} \epsilon_e$, the minimum injected Lorentz factor, the corresponding synchrotron frequency is
\begin{align}
\nu_m &= \frac{e B \gamma^2_m}{2 \pi m_e c} \propto t^{-2\alpha}\\ \nonumber
&= (1.08 \times 10^{11}~{\rm Hz}) \left(\frac{F_{\nu_p}}{\rm mJy}\right)^{-0.11} \left(\frac{D_L}{100~{\rm Mpc}}\right)^{-0.22} \left(\frac{\epsilon_e}{\epsilon_B}\right)^{-0.22} \epsilon^{2}_e f^{-0.22} \left(\frac{\nu_p}{\rm GHz}\right) \left(\frac{t}{t_p}\right)^{-2\alpha},
\end{align}
while for $\gamma_e = \gamma_c = \displaystyle{\frac{6 \pi m_e c}{\sigma_T B^2 t}}$, the critical Lorentz factor above which synchrotron cooling dominates and where $\sigma_T = 6.65 \times 10^{-25}$ cm$^2$ is the Thomson cross section, the corresponding synchrotron cooling frequency is
\begin{align}
\nu_c &= \frac{18 \pi m_e c e}{\sigma^2_T B^3 t^2} \propto t^{6 \alpha-2} \\ \nonumber
&= (7.27 \times 10^{17}~{\rm Hz}) \left(\frac{F_{\nu_p}}{\rm mJy}\right)^{0.33} \left(\frac{D_L}{100~{\rm Mpc}}\right)^{0.66} \left(\frac{\epsilon_e}{\epsilon_B}\right)^{0.66} f^{0.66} \left(\frac{\nu_p}{\rm GHz}\right)^{-3} \left(\frac{t}{t_p}\right)^{6\alpha-2} \left(\frac{t_p}{\rm day}\right)^{-2}.
\end{align}  

\section{Accretion}\label{fbacc}

Post the TDE,  most gravitationally bound stellar material returns to the pericenter when orbiting the SMBH. The fallback timescale is \citep{2011MNRAS.410..359L}
\begin{equation}
t_{\rm fb} = (41~{\rm days})~M^{1/2}_6 m^{(1-3 \xi)/2}_\star \beta^{-3},
\end{equation}
assuming that the main-sequence mass-radius relationship $R_\star = R_\odot m^{1-\xi}_\star$ \citep{1994sse..book.....K} is valid for the disrupted star with a scaled mass $m_\star = M_\star/M_\odot$ and radius $R_\star$, where the index $\xi \approx 0.2$ for $0.1 \leq m_\star \leq 1.0$, and $\xi \approx 0.4$ for $m_\star \geq 1.0$ \citep{2012ApJ...749...92K}. The other scaled quantities include $M_6 = M_\bullet/(10^6 M_\odot)$ and a penetration factor $\beta \equiv R_t/R_p$ defined as the ratio of the tidal radius of the SMBH $R_t$ to the pericenter distance $R_p$. 

The tidal radius of the SMBH $R_t$ can be expressed in terms of the SMBH mass and stellar properties \citep{1989ApJ...346L..13E} as
\begin{equation}\label{rprs}
R_t \approx \left(\frac{M_{\bullet}}{M_\star}\right)^{1/3} R_\star = (6.95\times 10^{12}~{\rm cm})~M^{1/3}_6 m^{(2/3)-\xi}_\star,
\end{equation}
which is then scaled in terms of the Schwarzschild radius $R_S = 2 G M_\bullet/c^2$ $=$ $(2.95 \times 10^{11}~{\rm cm})~M_6$, as
\begin{equation}
(R_t/R_S) = (23.58)~M^{-2/3}_6 m^{(2/3)-\xi}_\star.
\end{equation}
The distance of the closest approach to the SMBH is the pericenter distance $R_p \geq 3 R_S$, the innermost stable circular orbit for a non-rotating Schwarzschild black hole. With $R_t \geq R_p$, $R_t/R_S \geq R_p/R_S \geq 3$, the penetration factor $\beta \equiv R_t/R_p$ is in the range $1 \leq \beta \leq (7.86)~M^{-2/3}_6 m^{(2/3)-\xi}_\star$.

The fallback accretion rate onto the SMBH \citep{1989ApJ...346L..13E} is
\begin{equation}\label{Mdotfb}
\dot{M}_{\rm fb} = \frac{1}{3} \frac{M_\star}{t_{\rm fb}} \left(\frac{t}{t_{\rm fb}}\right)^{-5/3},
\end{equation}
at a time $t$ post the tidal disruption. The peak rate corresponds to $t = t_{\rm fb}$ where
\begin{equation}\label{Mdotp}
\dot{M}_{\rm p} = \dot{M}_{\rm fb}|_{t = t_{\rm fb}} \approx (3.0~M_\odot~{\rm yr}^{-1})~M^{-1/2}_6 m^{(1+3 \xi)/2}_\star \beta^3.
\end{equation}

The Eddington accretion rate is
\begin{equation}
\dot{M}_{\rm Edd} = \frac{L_{\rm Edd}}{\epsilon c^2} = \frac{4 \pi G M_\bullet m_p}{\epsilon c \sigma_T} = (0.023~{M_\odot}{\rm yr^{-1}})~M_6,
\end{equation}
for an assumed efficiency factor which we fix at a fiducial value $\epsilon = 0.1$ and where $\sigma_T = 6.65 \times 10^{-25}$ cm$^2$ is the Thomson cross section for electron scattering. Using the scaling  $t_d = t/(1~{\rm day})$, the fallback accretion rate and peak rate are super Eddington (see eqns. \ref{Mdotfb} and Eq. \ref{Mdotp}), with 
\begin{align}
\dot{m}_{\rm fb} &= (\dot{M}_{\rm fb}/\dot{M}_{\rm Edd}) = (6.34 \times 10^4)~M^{-2/3}_6 m^{(4-3 \xi)/3}_\star \beta^{-2} t^{-5/3}_d \label{mdotfb}\\
\dot{m}_{\rm p} &= (\dot{M}_{\rm p}/\dot{M}_{\rm Edd}) = (130.43)~M^{-3/2}_6 m^{(1+3 \xi)/2}_\star \beta^3. \label{mdotfbp}
\end{align}

\section{Outflow and photospheric properties}\label{outflowprop}

During the super-Eddington accretion phase, the returning tidal stream post the stellar disruption can encounter the stellar debris being accreted onto the SMBH, resulting in a shocked debris and the subsequent launching of outflows \cite[e.g.][]{2016ApJ...830..125J,2020MNRAS.492..686L,2020ApJ...902..108M}. The outflow is expected to be launched close to the circularization radius of the accreted material $R_C \approx 2 R_p = 2 R_t \beta^{-1}$ with an initial velocity \cite[e.g.][]{2009MNRAS.400.2070S},
\begin{equation}\label{vw}
v_{w} \approx \left(\frac{G M_\bullet}{R_C}\right)^{1/2} = (0.1~c)~M^{1/3}_6 m^{-(1/3-\xi/2)}_\star \beta^{1/2},
\end{equation}
with the equality corresponding to the escape velocity at the radius $R_C$. Assuming that the efficiency for mass-energy conversion during circularization is 1 \%, the associated timescale is \citep{2019ApJ...886..114H} 
\begin{equation}\label{tcirc}
t_{\rm circ} = (1.8 \times 10^{6}~{\rm s}) \beta^{-3/2} M^{-1/2}_6 m^{3 (1-2 \xi)/2}.
\end{equation}

The fallback accretion rate declines as $t^{-5/3}$ (see Eq. \ref{mdotfb}) with $\dot{M}_{\rm fb} = \dot{M}_{\rm Edd}$ denoting a transition from the super-Eddington to the sub-Eddington regime. The corresponding timescale $t_j$ is taken to be the lifetime of the outflow and is given by,
\begin{equation}\label{tj}
t_j	= (6.54 \times 10^{7}~s)~M^{-2/5}_6 m^{(4-3\xi)/5}_\star \beta^{-6/5}.
\end{equation}

With the above outflow velocity $v_w$ and $n \propto (R/R_{eq})^{-\gamma} = n_{eq} (t/t_p)^{-\gamma \alpha}$ to represent the number density decline with distance $R = R_{eq} (t/t_p)^\alpha$, the mass outflow rate is
\begin{align}
&\dot{M}_{w} = 4 \pi R^2 v_{w} n m_p \\
&= (1.18 \times 10^{-3}~M_\odot~{\rm yr}^{-1})~M^{1/3}_6 m^{-(1/3-\xi/2)}_\star \beta^{1/2} \left(\frac{F_{\nu_p}}{\rm mJy}\right)^{0.73} \left(\frac{D_L}{100~{\rm Mpc}}\right)^{1.46} \left(\frac{\epsilon_e}{\epsilon_B}\right)^{0.46} f^{-0.54} \left(\frac{t}{t_p}\right)^{\alpha (2-\gamma)}.
\end{align}

An upper limit on the particle number density in the vicinity of the accretion disk is based on the unbound stellar debris that is spatially and kinematically distributed in this region \citep{2009MNRAS.400.2070S}
\begin{equation}\label{naeqn}
n_{a} \leq (1.46 \times 10^{18}~{\rm cm^{-3}}) M^{1/6}_6 m^{(5/6)-(3 \xi/2)}_\star t^{-3}_d.
\end{equation}

Assuming that the constituent gas in the outflow is in thermal equilibrium with a consequent black-body spectrum, the total luminosity is approximated by the Stefan-Boltzmann law \citep{2009MNRAS.400.2070S} such that  
\begin{equation}\label{Ljeq}
L_{j} = \sigma_{\rm SB} (4 \pi R^2_{\rm ph}) T^{4}_{\rm ph} = (6.68 \times 10^{44}~{\rm erg~s^{-1}})~M^{8/9}_6 m^{(2-\xi)/6}_\star \beta^{-2/3} t^{-5/9}_d,
\end{equation}
where $\sigma_{\rm SB} = 5.67 \times 10^{-5}~{\rm erg~cm^{-2}~s^{-1}~K^{-4}}$ is the Stefan-Boltzmann constant, a photospheric radius $R_{\rm ph}$ and an associated thermal temperature $T_{\rm ph}$,
\begin{align}
R_{\rm ph} &= (6.3 \times 10^{16}~{\rm cm})~m^{5/3-3\xi/2}_\star \beta^{-2} t^{-5/3}_d \label{Rph} \\
T_{\rm ph} &= (1.25\times 10^{3}~{\rm K})~M^{2/9}_6 m^{3/4-17\xi/24}_\star \beta^{5/6} t^{25/36}_d, \label{Tph}
\end{align}
assuming a mass outflow rate $\dot{M}_w \leq 0.1 \dot{M}_{\rm fb}$ \cite[e.g.][]{2015ApJ...814..141M} and the equality in Eq. \ref{vw}. Also employed in the above equation are the expressions for $R_p/(3 R_S)$ and $\dot{m}_{\rm fb}$ from Eq. \ref{rprs} and Eq. \ref{mdotfb}, respectively. The corresponding total kinetic energy $E_j$ is estimated using the above thermal luminosity and an additive component from the power-law tail during the declining phase, 
\begin{equation}\label{Ejeq}
E_j	= \int^{t_j}_{0} L_j dt + \int^{t}_{t_j} L_j (t/t_j)^{-5/3} dt \approx (3.81 \times 10^{51}~{\rm erg})~M^{32/45}_6 m^{(62-39 \xi)/90}_\star \beta^{-6/5}.
\end{equation}

The number density of material in the outflow is given by
\begin{equation}
n_j = \frac{L_j}{4 \pi R^2 m_p c^3}.
\end{equation}
Using $L_j$ from Eq. \ref{Ljeq} and the scalings $n = n_{eq} (t/t_p)^{-\gamma \alpha}$ and $R = R_{eq} (t/t_p)^\alpha$, the density contrast between the CNM and the outflow,
\begin{equation}\label{kappaeqn}
\kappa = \frac{n}{n_j} \approx (1.0)~M^{-8/9}_6 m^{-(2-\xi)/6}_\star \beta^{2/3} \left(\frac{F_{\nu_p}}{\rm mJy}\right)^{0.73} \left(\frac{D_L}{100~{\rm Mpc}}\right)^{1.46} \left(\frac{\epsilon_e}{\epsilon_B}\right)^{0.46} f^{-0.54} \left(\frac{t}{t_p}\right)^{(2-\gamma) \alpha+5/9} \left(\frac{t_p}{\rm day}\right)^{5/9}.
\end{equation}

The maximum energy to which protons can be accelerated by the outflow is limited by comparing the associated timescale with the dynamical timescale \cite[e.g.][]{2017PhRvD..96f3007Z}. Using the scalings for the size, velocity and magnetic field strength in the emission region, the maximum energy is 
\begin{align}
E_p &\leqslant \frac{3 v^2}{20 c^2} e B c t = \frac{3}{20} \frac{e \alpha^2}{c t} R^{2}_{eq} B_{eq} \label{Epoutflow}\\ \nonumber
 &\leqslant (5.23 \times 10^{18}~{\rm eV}) \left(\frac{F_{\nu_p}}{\rm mJy}\right)^{0.84} \left(\frac{D_L}{100~{\rm Mpc}}\right)^{1.67} \left(\frac{\epsilon_e}{\epsilon_B}\right)^{-0.33} f^{-0.33} \left(\frac{\nu_p}{\rm GHz}\right)^{-1} \left(\frac{t}{\rm day}\right)^{-1} 
\end{align}

The maximum energy to which protons can be accelerated during the radiatively inefficient accretion flow is limited by comparing the timescale for plasma turbulence acceleration with the diffusion timescale \citep{2019ApJ...886..114H}. Using the photospheric radius $R = R_{ph}$, and assuming a plasma beta of 3, and a proton number density equal to the particle number density calculated above, the maximum energy of protons is \citep{2019ApJ...886..114H},
\begin{align}
E_p &\leqslant \frac{2 \pi^{1/2} e}{3} \left(\frac{R}{R_S}\right)^{-1} R_S (m_p n_p c^2)^{1/2}\label{EpRIAF} \\ \nonumber
E_p &\leqslant (2.33 \times 10^{16}~{\rm eV}) M^{25/12}_6 m^{-(5/4)-(9 \xi/8)}_\star \beta^2 \left(\frac{t}{\rm day}\right)^{1/6}. 
\end{align}



\end{appendix}

\bibliography{ref}

\begin{thebibliography}{}
\expandafter\ifx\csname natexlab\endcsname\relax\def\natexlab#1{#1}\fi
\providecommand{\url}[1]{\href{#1}{#1}}
\providecommand{\dodoi}[1]{doi:~\href{http://doi.org/#1}{\nolinkurl{#1}}}
\providecommand{\doeprint}[1]{\href{http://ascl.net/#1}{\nolinkurl{http://ascl.net/#1}}}
\providecommand{\doarXiv}[1]{\href{https://arxiv.org/abs/#1}{\nolinkurl{https://arxiv.org/abs/#1}}}

\bibitem[{{Alexander} {et~al.}(2016){Alexander}, {Berger}, {Guillochon},
  {Zauderer}, \& {Williams}}]{2016ApJ...819L..25A}
{Alexander}, K.~D., {Berger}, E., {Guillochon}, J., {Zauderer}, B.~A., \&
  {Williams}, P.~K.~G. 2016, \apjl, 819, L25

\bibitem[{{Alexander} {et~al.}(2020){Alexander}, {van Velzen}, {Horesh}, \&
  {Zauderer}}]{2020SSRv..216...81A}
{Alexander}, K.~D., {van Velzen}, S., {Horesh}, A., \& {Zauderer}, B.~A. 2020,
  \ssr, 216, 81

\bibitem[{{Anderson} {et~al.}(2020){Anderson}, {Mooley}, {Hallinan}, {Dong},
  {Phinney}, {Horesh}, {Bourke}, {Cenko}, {Frail}, {Kulkarni}, \&
  {Myers}}]{2020ApJ...903..116A}
{Anderson}, M.~M., {Mooley}, K.~P., {Hallinan}, G., {et~al.} 2020, \apj, 903,
  116

\bibitem[{{Beasley} \& {Conway}(1995)}]{1995ASPC...82..327B}
{Beasley}, A.~J., \& {Conway}, J.~E. 1995, in Astronomical Society of the
  Pacific Conference Series, Vol.~82, Very Long Baseline Interferometry and the
  VLBA, ed. J.~A. {Zensus}, P.~J. {Diamond}, \& P.~J. {Napier}, 327

\bibitem[{{Berger} {et~al.}(2012){Berger}, {Zauderer}, {Pooley}, {Soderberg},
  {Sari}, {Brunthaler}, \& {Bietenholz}}]{2012ApJ...748...36B}
{Berger}, E., {Zauderer}, A., {Pooley}, G.~G., {et~al.} 2012, \apj, 748, 36,
  \dodoi{10.1088/0004-637X/748/1/36}

\bibitem[{{Biehl} {et~al.}(2018){Biehl}, {Boncioli}, {Lunardini}, \&
  {Winter}}]{2018NatSR...810828B}
{Biehl}, D., {Boncioli}, D., {Lunardini}, C., \& {Winter}, W. 2018, Scientific
  Reports, 8, 10828, \dodoi{10.1038/s41598-018-29022-4}

\bibitem[{{Blandford} \& {McKee}(1976)}]{1976PhFl...19.1130B}
{Blandford}, R.~D., \& {McKee}, C.~F. 1976, Physics of Fluids, 19, 1130

\bibitem[{{Bright} {et~al.}(2018){Bright}, {Fender}, {Motta}, {Mooley},
  {Perrott}, {van Velzen}, {Carey}, {Hickish}, {Razavi-Ghods}, {Titterington},
  {Scott}, {Grainge}, {Scaife}, {Cantwell}, \& {Rumsey}}]{2018MNRAS.475.4011B}
{Bright}, J.~S., {Fender}, R.~P., {Motta}, S.~E., {et~al.} 2018, \mnras, 475,
  4011, \dodoi{10.1093/mnras/sty077}

\bibitem[{{Cannizzaro} {et~al.}(2021){Cannizzaro}, {Wevers}, {Jonker},
  {P{\'e}rez-Torres}, {Moldon}, {Mata-S{\'a}nchez}, {Leloudas}, {Pasham},
  {Mattila}, {Arcavi}, {Decker French}, {Onori}, {Inserra}, {Nicholl},
  {Gromadzki}, {Chen}, {M{\"u}ller-Bravo}, {Short}, {Anderson}, {Young},
  {Gendreau}, {Arzoumanian}, {L{\"o}wenstein}, {Remillard}, {Roy}, \&
  {Hiramatsu}}]{2021MNRAS.504..792C}
{Cannizzaro}, G., {Wevers}, T., {Jonker}, P.~G., {et~al.} 2021, \mnras, 504,
  792, \dodoi{10.1093/mnras/stab851}

\bibitem[{{Cendes} {et~al.}(2021{\natexlab{a}}){Cendes}, {Alexander}, {Berger},
  {Eftekhari}, {Williams}, \& {Chornock}}]{2021arXiv210306299C}
{Cendes}, Y., {Alexander}, K.~D., {Berger}, E., {et~al.} 2021{\natexlab{a}},
  arXiv e-prints, arXiv:2103.06299.
\newblock \doarXiv{2103.06299}

\bibitem[{{Cendes} {et~al.}(2021{\natexlab{b}}){Cendes}, {Eftekhari}, {Berger},
  \& {Polisensky}}]{2021ApJ...908..125C}
{Cendes}, Y., {Eftekhari}, T., {Berger}, E., \& {Polisensky}, E.
  2021{\natexlab{b}}, \apj, 908, 125, \dodoi{10.3847/1538-4357/abd323}

\bibitem[{{Cheng} {et~al.}(2020){Cheng}, {An}, {Frey}, {Hong}, {He},
  {Kellermann}, {Lister}, {Lao}, {Li}, {Mohan}, {Yang}, {Wu}, {Zhang}, {Zhang},
  \& {Zhao}}]{2020ApJS..247...57C}
{Cheng}, X.~P., {An}, T., {Frey}, S., {et~al.} 2020, \apjs, 247, 57,
  \dodoi{10.3847/1538-4365/ab791f}

\bibitem[{{Chevalier}(1998)}]{1998ApJ...499..810C}
{Chevalier}, R.~A. 1998, \apj, 499, 810, \dodoi{10.1086/305676}

\bibitem[{{Eftekhari} {et~al.}(2018){Eftekhari}, {Berger}, {Zauderer},
  {Margutti}, \& {Alexander}}]{2018ApJ...854...86E}
{Eftekhari}, T., {Berger}, E., {Zauderer}, B.~A., {Margutti}, R., \&
  {Alexander}, K.~D. 2018, \apj, 854, 86, \dodoi{10.3847/1538-4357/aaa8e0}

\bibitem[{{Evans} \& {Kochanek}(1989)}]{1989ApJ...346L..13E}
{Evans}, C.~R., \& {Kochanek}, C.~S. 1989, \apjl, 346, L13

\bibitem[{{Farrar} \& {Gruzinov}(2009)}]{2009ApJ...693..329F}
{Farrar}, G.~R., \& {Gruzinov}, A. 2009, \apj, 693, 329,
  \dodoi{10.1088/0004-637X/693/1/329}

\bibitem[{{Fomalont}(1999)}]{1999ASPC..180..301F}
{Fomalont}, E.~B. 1999, in Astronomical Society of the Pacific Conference
  Series, Vol. 180, Synthesis Imaging in Radio Astronomy II, ed. G.~B.
  {Taylor}, C.~L. {Carilli}, \& R.~A. {Perley}, 301

\bibitem[{{Ge} {et~al.}(2021){Ge}, {Liu}, {Niu}, {Chen}, \&
  {Wang}}]{2021Innov...200118G}
{Ge}, C., {Liu}, R.-Y., {Niu}, S., {Chen}, Y., \& {Wang}, X.-Y. 2021, The
  Innovation, 2, 100118, \dodoi{10.1016/j.xinn.2021.100118}

\bibitem[{{Giommi} {et~al.}(2020){Giommi}, {Glauch}, {Padovani}, {Resconi},
  {Turcati}, \& {Chang}}]{2020MNRAS.497..865G}
{Giommi}, P., {Glauch}, T., {Padovani}, P., {et~al.} 2020, \mnras, 497, 865

\bibitem[{{Greisen}(2003)}]{2003ASSL..285..109G}
{Greisen}, E.~W. 2003, {AIPS, the VLA, and the VLBA}, ed. A.~{Heck}, Vol. 285,
  109, \dodoi{10.1007/0-306-48080-8_7}

\bibitem[{{Gu{\'e}pin} \& {Kotera}(2017)}]{2017A&A...603A..76G}
{Gu{\'e}pin}, C., \& {Kotera}, K. 2017, \aap, 603, A76

\bibitem[{{Gu{\'e}pin} {et~al.}(2018){Gu{\'e}pin}, {Kotera}, {Barausse},
  {Fang}, \& {Murase}}]{2018A&A...616A.179G}
{Gu{\'e}pin}, C., {Kotera}, K., {Barausse}, E., {Fang}, K., \& {Murase}, K.
  2018, \aap, 616, A179, \dodoi{10.1051/0004-6361/201732392}

\bibitem[{{Hayasaki} \& {Yamazaki}(2019)}]{2019ApJ...886..114H}
{Hayasaki}, K., \& {Yamazaki}, R. 2019, \apj, 886, 114,
  \dodoi{10.3847/1538-4357/ab44ca}

\bibitem[{{HESS Collaboration} {et~al.}(2016){HESS Collaboration},
  {Abramowski}, {Aharonian}, {Benkhali}, {Akhperjanian}, {Ang{\"u}ner},
  {Backes}, {Balzer}, {Becherini}, {Tjus}, {Berge}, {Bernhard}, {Bernl{\"o}hr},
  {Birsin}, {Blackwell}, {B{\"o}ttcher}, {Boisson}, {Bolmont}, {Bordas},
  {Bregeon}, {Brun}, {Brun}, {Bryan}, {Bulik}, {Carr}, {Casanova},
  {Chakraborty}, {Chalme-Calvet}, {Chaves}, {Chen}, {Chr{\'e}tien},
  {Colafrancesco}, {Cologna}, {Conrad}, {Couturier}, {Cui}, {Davids},
  {Degrange}, {Deil}, {Dewilt}, {Djannati-Ata{\"\i}}, {Domainko}, {Donath},
  {Drury}, {Dubus}, {Dutson}, {Dyks}, {Dyrda}, {Edwards}, {Egberts}, {Eger},
  {Ernenwein}, {Espigat}, {Farnier}, {Fegan}, {Feinstein}, {Fernandes},
  {Fernandez}, {Fiasson}, {Fontaine}, {F{\"o}rster}, {F{\"u}{\ss}ling},
  {Gabici}, {Gajdus}, {Gallant}, {Garrigoux}, {Giavitto}, {Giebels},
  {Glicenstein}, {Gottschall}, {Goyal}, {Grondin}, {Grudzi{\'n}ska}, {Hadasch},
  {H{\"a}ffner}, {Hahn}, {Hawkes}, {Heinzelmann}, {Henri}, {Hermann}, {Hervet},
  {Hillert}, {Hinton}, {Hofmann}, {Hofverberg}, {Hoischen}, {Holler}, {Horns},
  {Ivascenko}, {Jacholkowska}, {Jamrozy}, {Janiak}, {Jankowsky},
  {Jung-Richardt}, {Kastendieck}, {Katarzy{\'n}ski}, {Katz}, {Kerszberg},
  {Kh{\'e}lifi}, {Kieffer}, {Klepser}, {Klochkov}, {Klu{\'z}niak}, {Kolitzus},
  {Komin}, {Kosack}, {Krakau}, {Krayzel}, {Kr{\"u}ger}, {Laffon}, {Lamanna},
  {Lau}, {Lefaucheur}, {Lefranc}, {Lemi{\'e}re}, {Lemoine-Goumard}, {Lenain},
  {Lohse}, {Lopatin}, {Lu}, {Lui}, {Marandon}, {Marcowith}, {Mariaud}, {Marx},
  {Maurin}, {Maxted}, {Mayer}, {Meintjes}, {Menzler}, {Meyer}, {Mitchell},
  {Moderski}, {Mohamed}, {Mor{\r{a}}}, {Moulin}, {Murach}, {de Naurois},
  {Niemiec}, {Oakes}, {Odaka}, {{\"O}ttl}, {Ohm}, {Opitz}, {Ostrowski}, {Oya},
  {Panter}, {Parsons}, {Arribas}, {Pekeur}, {Pelletier}, {Petrucci}, {Peyaud},
  {Pita}, {Poon}, {Prokoph}, {P{\"u}hlhofer}, {Punch}, {Quirrenbach}, {Raab},
  {Reichardt}, {Reimer}, {Reimer}, {Renaud}, {de Los Reyes}, {Rieger},
  {Romoli}, {Rosier-Lees}, {Rowell}, {Rudak}, {Rulten}, {Sahakian}, {Salek},
  {Sanchez}, {Santangelo}, {Sasaki}, {Schlickeiser}, {Sch{\"u}ssler}, {Schulz},
  {Schwanke}, {Schwemmer}, {Seyffert}, {Simoni}, {Sol}, {Spanier}, {Spengler},
  {Spies}, {Stawarz}, {Steenkamp}, {Stegmann}, {Stinzing}, {Stycz}, {Sushch},
  {Tavernet}, {Tavernier}, {Taylor}, {Terrier}, {Tluczykont}, {Trichard},
  {Tuffs}, {Valerius}, {van der Walt}, {van Eldik}, {van Soelen},
  {Vasileiadis}, {Veh}, {Venter}, {Viana}, {Vincent}, {Vink}, {Voisin},
  {V{\"o}lk}, {Vuillaume}, {Wagner}, {Wagner}, {Wagner}, {Weidinger},
  {Weitzel}, {White}, {Wierzcholska}, {Willmann}, {W{\"o}rnlein}, {Wouters},
  {Yang}, {Zabalza}, {Zaborov}, {Zacharias}, {Zdziarski}, {Zech}, {Zefi}, \&
  {{\.Z}ywucka}}]{2016Natur.531..476H}
{HESS Collaboration}, {Abramowski}, A., {Aharonian}, F., {et~al.} 2016, \nat,
  531, 476, \dodoi{10.1038/nature17147}

\bibitem[{{Ho} {et~al.}(2019){Ho}, {Phinney}, {Ravi}, {Kulkarni}, {Petitpas},
  {Emonts}, {Bhalerao}, {Blundell}, {Cenko}, {Dobie}, {Howie}, {Kamraj},
  {Kasliwal}, {Murphy}, {Perley}, {Sridharan}, \& {Yoon}}]{2019ApJ...871...73H}
{Ho}, A. Y.~Q., {Phinney}, E.~S., {Ravi}, V., {et~al.} 2019, \apj, 871, 73,
  \dodoi{10.3847/1538-4357/aaf473}

\bibitem[{{Hooper} {et~al.}(2019){Hooper}, {Linden}, \&
  {Vieregg}}]{2019JCAP...02..012H}
{Hooper}, D., {Linden}, T., \& {Vieregg}, A. 2019, \jcap, 2019, 012,
  \dodoi{10.1088/1475-7516/2019/02/012}

\bibitem[{{Horesh} {et~al.}(2021){Horesh}, {Cenko}, \&
  {Arcavi}}]{2021NatAs...5..491H}
{Horesh}, A., {Cenko}, S.~B., \& {Arcavi}, I. 2021, Nature Astronomy, 5, 491,
  \dodoi{10.1038/s41550-021-01300-8}

\bibitem[{{IceCube Collaboration}(2019)}]{2019GCN.25913....1I}
{IceCube Collaboration}. 2019, GRB Coordinates Network, 25913, 1

\bibitem[{{Jiang} {et~al.}(2016){Jiang}, {Guillochon}, \&
  {Loeb}}]{2016ApJ...830..125J}
{Jiang}, Y.-F., {Guillochon}, J., \& {Loeb}, A. 2016, \apj, 830, 125

\bibitem[{{Keimpema} {et~al.}(2015){Keimpema}, {Kettenis}, {Pogrebenko},
  {Campbell}, {Cim{\'o}}, {Duev}, {Eldering}, {Kruithof}, {van Langevelde},
  {Marchal}, {Molera Calv{\'e}s}, {Ozdemir}, {Paragi}, {Pidopryhora},
  {Szomoru}, \& {Yang}}]{2015ExA....39..259K}
{Keimpema}, A., {Kettenis}, M.~M., {Pogrebenko}, S.~V., {et~al.} 2015,
  Experimental Astronomy, 39, 259

\bibitem[{{Kippenhahn} \& {Weigert}(1994)}]{1994sse..book.....K}
{Kippenhahn}, R., \& {Weigert}, A. 1994, {Stellar Structure and Evolution}

\bibitem[{{Krolik} {et~al.}(2016){Krolik}, {Piran}, {Svirski}, \&
  {Cheng}}]{2016ApJ...827..127K}
{Krolik}, J., {Piran}, T., {Svirski}, G., \& {Cheng}, R.~M. 2016, \apj, 827,
  127, \dodoi{10.3847/0004-637X/827/2/127}

\bibitem[{{Krolik} \& {Piran}(2012)}]{2012ApJ...749...92K}
{Krolik}, J.~H., \& {Piran}, T. 2012, \apj, 749, 92

\bibitem[{{Lee} {et~al.}(2020){Lee}, {Hung}, {Matheson}, {Soraisam}, {Narayan},
  {Saha}, {Stubens}, \& {Wolf}}]{2020ApJ...892L...1L}
{Lee}, C.-H., {Hung}, T., {Matheson}, T., {et~al.} 2020, \apjl, 892, L1

\bibitem[{{Levan} {et~al.}(2011){Levan}, {Tanvir}, {Cenko}, {Perley},
  {Wiersema}, {Bloom}, {Fruchter}, {de Ugarte Postigo}, {O'Brien}, {Butler},
  {van der Horst}, {Leloudas}, {Morgan}, {Misra}, {Bower}, {Farihi},
  {Tunnicliffe}, {Modjaz}, {Silverman}, {Hjorth}, {Th{\"o}ne}, {Cucchiara},
  {Cer{\'o}n}, {Castro-Tirado}, {Arnold}, {Bremer}, {Brodie}, {Carroll},
  {Cooper}, {Curran}, {Cutri}, {Ehle}, {Forbes}, {Fynbo}, {Gorosabel},
  {Graham}, {Hoffman}, {Guziy}, {Jakobsson}, {Kamble}, {Kerr}, {Kasliwal},
  {Kouveliotou}, {Kocevski}, {Law}, {Nugent}, {Ofek}, {Poznanski}, {Quimby},
  {Rol}, {Romanowsky}, {S{\'a}nchez-Ram{\'\i}rez}, {Schulze}, {Singh}, {van
  Spaandonk}, {Starling}, {Strom}, {Tello}, {Vaduvescu}, {Wheatley}, {Wijers},
  {Winters}, \& {Xu}}]{2011Sci...333..199L}
{Levan}, A.~J., {Tanvir}, N.~R., {Cenko}, S.~B., {et~al.} 2011, Science, 333,
  199

\bibitem[{{Liu} {et~al.}(2020){Liu}, {Xi}, \& {Wang}}]{2020PhRvD.102h3028L}
{Liu}, R.-Y., {Xi}, S.-Q., \& {Wang}, X.-Y. 2020, \prd, 102, 083028

\bibitem[{{Lodato} \& {Rossi}(2011)}]{2011MNRAS.410..359L}
{Lodato}, G., \& {Rossi}, E.~M. 2011, \mnras, 410, 359

\bibitem[{{Lu} \& {Bonnerot}(2020)}]{2020MNRAS.492..686L}
{Lu}, W., \& {Bonnerot}, C. 2020, \mnras, 492, 686

\bibitem[{{Mageshwaran} \& {Mangalam}(2015)}]{2015ApJ...814..141M}
{Mageshwaran}, T., \& {Mangalam}, A. 2015, \apj, 814, 141

\bibitem[{{Mannheim}(1993)}]{1993A&A...269...67M}
{Mannheim}, K. 1993, \aap, 269, 67

\bibitem[{{Matsumoto} {et~al.}(2021){Matsumoto}, {Piran}, \&
  {Krolik}}]{2021arXiv210902648M}
{Matsumoto}, T., {Piran}, T., \& {Krolik}, J.~H. 2021, arXiv e-prints,
  arXiv:2109.02648.
\newblock \doarXiv{2109.02648}

\bibitem[{{Mattila} {et~al.}(2018){Mattila}, {P{\'e}rez-Torres}, {Efstathiou},
  {Mimica}, {Fraser}, {Kankare}, {Alberdi}, {Aloy}, {Heikkil{\"a}}, {Jonker},
  {Lundqvist}, {Mart{\'\i}-Vidal}, {Meikle}, {Romero-Ca{\~n}izales}, {Smartt},
  {Tsygankov}, {Varenius}, {Alonso-Herrero}, {Bondi}, {Fransson},
  {Herrero-Illana}, {Kangas}, {Kotak}, {Ram{\'\i}rez-Olivencia},
  {V{\"a}is{\"a}nen}, {Beswick}, {Clements}, {Greimel}, {Harmanen},
  {Kotilainen}, {Nandra}, {Reynolds}, {Ryder}, {Walton}, {Wiik}, \&
  {{\"O}stlin}}]{2018Sci...361..482M}
{Mattila}, S., {P{\'e}rez-Torres}, M., {Efstathiou}, A., {et~al.} 2018,
  Science, 361, 482

\bibitem[{{Metzger} {et~al.}(2012){Metzger}, {Giannios}, \&
  {Mimica}}]{2012MNRAS.420.3528M}
{Metzger}, B.~D., {Giannios}, D., \& {Mimica}, P. 2012, \mnras, 420, 3528

\bibitem[{{Mohan} {et~al.}(2020){Mohan}, {An}, \& {Yang}}]{2020ApJ...888L..24M}
{Mohan}, P., {An}, T., \& {Yang}, J. 2020, \apjl, 888, L24

\bibitem[{{Mohan} \& {Mangalam}(2015)}]{2015ApJ...805...91M}
{Mohan}, P., \& {Mangalam}, A. 2015, \apj, 805, 91,
  \dodoi{10.1088/0004-637X/805/2/91}

\bibitem[{{Murase} {et~al.}(2020){Murase}, {Kimura}, {Zhang}, {Oikonomou}, \&
  {Petropoulou}}]{2020ApJ...902..108M}
{Murase}, K., {Kimura}, S.~S., {Zhang}, B.~T., {Oikonomou}, F., \&
  {Petropoulou}, M. 2020, \apj, 902, 108

\bibitem[{{Murase} \& {Waxman}(2016)}]{2016PhRvD..94j3006M}
{Murase}, K., \& {Waxman}, E. 2016, \prd, 94, 103006,
  \dodoi{10.1103/PhysRevD.94.103006}

\bibitem[{{Nicholl} {et~al.}(2019){Nicholl}, {Short}, {Angus}, {Muller},
  {Pursiainen}, {Barbarino}, {Dennefeld}, {Williams}, {Perley}, {Benetti},
  {Anderson}, {Chen}, {Inserra}, {Yaron}, {Young}, {Manulis}, {Tonry},
  {Denneau}, {Heinze}, {Weiland}, {Stalder}, {Rest}, {Smith}, {Smartt},
  {McBrien}, \& {Srivastav}}]{2019ATel12752....1N}
{Nicholl}, M., {Short}, P., {Angus}, C., {et~al.} 2019, The Astronomer's
  Telegram, 12752, 1

\bibitem[{{Nordin} {et~al.}(2019){Nordin}, {Brinnel}, {Giomi}, {Santen},
  {Gal-Yam}, {Yaron}, \& {Schulze}}]{2019TNSTR.615....1N}
{Nordin}, J., {Brinnel}, V., {Giomi}, M., {et~al.} 2019, Transient Name Server
  Discovery Report, 2019-615, 1

\bibitem[{{Pacholczyk}(1970)}]{1970ranp.book.....P}
{Pacholczyk}, A.~G. 1970, {Radio astrophysics. Nonthermal processes in galactic
  and extragalactic sources}

\bibitem[{{Phinney}(1989)}]{1989IAUS..136..543P}
{Phinney}, E.~S. 1989, in The Center of the Galaxy, ed. M.~{Morris}, Vol. 136,
  543

\bibitem[{{Pradel} {et~al.}(2006){Pradel}, {Charlot}, \&
  {Lestrade}}]{2006A&A...452.1099P}
{Pradel}, N., {Charlot}, P., \& {Lestrade}, J.~F. 2006, \aap, 452, 1099.
\newblock \doarXiv{astro-ph/0603015}

\bibitem[{{Rachen} \& {M{\'e}sz{\'a}ros}(1998)}]{1998PhRvD..58l3005R}
{Rachen}, J.~P., \& {M{\'e}sz{\'a}ros}, P. 1998, \prd, 58, 123005,
  \dodoi{10.1103/PhysRevD.58.123005}

\bibitem[{{Rees}(1988)}]{1988Natur.333..523R}
{Rees}, M.~J. 1988, \nat, 333, 523

\bibitem[{{Romero-Ca{\~n}izales} {et~al.}(2016){Romero-Ca{\~n}izales},
  {Prieto}, {Chen}, {Kochanek}, {Dong}, {Holoien}, {Stanek}, \&
  {Liu}}]{2016ApJ...832L..10R}
{Romero-Ca{\~n}izales}, C., {Prieto}, J.~L., {Chen}, X., {et~al.} 2016, \apjl,
  832, L10

\bibitem[{{Roth} {et~al.}(2020){Roth}, {Rossi}, {Krolik}, {Piran}, {Mockler},
  \& {Kasen}}]{2020SSRv..216..114R}
{Roth}, N., {Rossi}, E.~M., {Krolik}, J., {et~al.} 2020, \ssr, 216, 114,
  \dodoi{10.1007/s11214-020-00735-1}

\bibitem[{{Saxton} {et~al.}(2020){Saxton}, {Komossa}, {Auchettl}, \&
  {Jonker}}]{2020SSRv..216...85S}
{Saxton}, R., {Komossa}, S., {Auchettl}, K., \& {Jonker}, P.~G. 2020, \ssr,
  216, 85, \dodoi{10.1007/s11214-020-00708-4}

\bibitem[{{Scott} \& {Readhead}(1977)}]{1977MNRAS.180..539S}
{Scott}, M.~A., \& {Readhead}, A.~C.~S. 1977, \mnras, 180, 539,
  \dodoi{10.1093/mnras/180.4.539}

\bibitem[{{Shepherd}(1997)}]{1997ASPC..125...77S}
{Shepherd}, M.~C. 1997, in Astronomical Society of the Pacific Conference
  Series, Vol. 125, Astronomical Data Analysis Software and Systems VI, ed.
  G.~{Hunt} \& H.~{Payne}, 77

\bibitem[{{Stein} {et~al.}(2021){Stein}, {Velzen}, {Kowalski}, {Franckowiak},
  {Gezari}, {Miller-Jones}, {Frederick}, {Sfaradi}, {Bietenholz}, {Horesh},
  {Fender}, {Garrappa}, {Ahumada}, {Andreoni}, {Belicki}, {Bellm},
  {B{\"o}ttcher}, {Brinnel}, {Burruss}, {Cenko}, {Coughlin}, {Cunningham},
  {Drake}, {Farrar}, {Feeney}, {Foley}, {Gal-Yam}, {Golkhou}, {Goobar},
  {Graham}, {Hammerstein}, {Helou}, {Hung}, {Kasliwal}, {Kilpatrick}, {Kong},
  {Kupfer}, {Laher}, {Mahabal}, {Masci}, {Necker}, {Nordin}, {Perley},
  {Rigault}, {Reusch}, {Rodriguez}, {Rojas-Bravo}, {Rusholme}, {Shupe},
  {Singer}, {Sollerman}, {Soumagnac}, {Stern}, {Taggart}, {van Santen}, {Ward},
  {Woudt}, \& {Yao}}]{2021NatAs...5..510S}
{Stein}, R., {Velzen}, S.~v., {Kowalski}, M., {et~al.} 2021, Nature Astronomy,
  5, 510, \dodoi{10.1038/s41550-020-01295-8}

\bibitem[{{Strubbe} \& {Quataert}(2009)}]{2009MNRAS.400.2070S}
{Strubbe}, L.~E., \& {Quataert}, E. 2009, \mnras, 400, 2070

\bibitem[{{van Velzen} {et~al.}(2016){van Velzen}, {Anderson}, {Stone},
  {Fraser}, {Wevers}, {Metzger}, {Jonker}, {van der Horst}, {Staley}, {Mendez},
  {Miller-Jones}, {Hodgkin}, {Campbell}, \& {Fender}}]{2016Sci...351...62V}
{van Velzen}, S., {Anderson}, G.~E., {Stone}, N.~C., {et~al.} 2016, Science,
  351, 62, \dodoi{10.1126/science.aad1182}

\bibitem[{{van Velzen} {et~al.}(2021){van Velzen}, {Gezari}, {Hammerstein},
  {Roth}, {Frederick}, {Ward}, {Hung}, {Cenko}, {Stein}, {Perley}, {Taggart},
  {Foley}, {Sollerman}, {Blagorodnova}, {Andreoni}, {Bellm}, {Brinnel}, {De},
  {Dekany}, {Feeney}, {Fremling}, {Giomi}, {Golkhou}, {Graham}, {Ho},
  {Kasliwal}, {Kilpatrick}, {Kulkarni}, {Kupfer}, {Laher}, {Mahabal}, {Masci},
  {Miller}, {Nordin}, {Riddle}, {Rusholme}, {van Santen}, {Sharma}, {Shupe}, \&
  {Soumagnac}}]{2021ApJ...908....4V}
{van Velzen}, S., {Gezari}, S., {Hammerstein}, E., {et~al.} 2021, \apj, 908, 4,
  \dodoi{10.3847/1538-4357/abc258}

\bibitem[{{Waxman} \& {Bahcall}(1999)}]{1999PhRvD..59b3002W}
{Waxman}, E., \& {Bahcall}, J. 1999, \prd, 59, 023002,
  \dodoi{10.1103/PhysRevD.59.023002}

\bibitem[{{Winter} \& {Lunardini}(2021)}]{2021NatAs...5..472W}
{Winter}, W., \& {Lunardini}, C. 2021, Nature Astronomy, 5, 472,
  \dodoi{10.1038/s41550-021-01305-3}

\bibitem[{{Wright}(2006)}]{2006PASP..118.1711W}
{Wright}, E.~L. 2006, \pasp, 118, 1711

\bibitem[{{Yang} {et~al.}(2021){Yang}, {Paragi}, {Nardini}, {Baan}, {Fan},
  {Mohan}, {Varenius}, \& {An}}]{2021MNRAS.500.2620Y}
{Yang}, J., {Paragi}, Z., {Nardini}, E., {et~al.} 2021, \mnras, 500, 2620,
  \dodoi{10.1093/mnras/staa2445}

\bibitem[{{Yang} {et~al.}(2016){Yang}, {Paragi}, {van der Horst}, {Gurvits},
  {Campbell}, {Giannios}, {An}, \& {Komossa}}]{2016MNRAS.462L..66Y}
{Yang}, J., {Paragi}, Z., {van der Horst}, A.~J., {et~al.} 2016, \mnras, 462,
  L66

\bibitem[{{Zauderer} {et~al.}(2013){Zauderer}, {Berger}, {Margutti}, {Pooley},
  {Sari}, {Soderberg}, {Brunthaler}, \& {Bietenholz}}]{2013ApJ...767..152Z}
{Zauderer}, B.~A., {Berger}, E., {Margutti}, R., {et~al.} 2013, \apj, 767, 152,
  \dodoi{10.1088/0004-637X/767/2/152}

\bibitem[{{Zauderer} {et~al.}(2011){Zauderer}, {Berger}, {Soderberg}, {Loeb},
  {Narayan}, {Frail}, {Petitpas}, {Brunthaler}, {Chornock}, {Carpenter},
  {Pooley}, {Mooley}, {Kulkarni}, {Margutti}, {Fox}, {Nakar}, {Patel},
  {Volgenau}, {Culverhouse}, {Bietenholz}, {Rupen}, {Max-Moerbeck}, {Readhead},
  {Richards}, {Shepherd}, {Storm}, \& {Hull}}]{2011Natur.476..425Z}
{Zauderer}, B.~A., {Berger}, E., {Soderberg}, A.~M., {et~al.} 2011, \nat, 476,
  425

\bibitem[{{Zhang} {et~al.}(2017){Zhang}, {Murase}, {Oikonomou}, \&
  {Li}}]{2017PhRvD..96f3007Z}
{Zhang}, B.~T., {Murase}, K., {Oikonomou}, F., \& {Li}, Z. 2017, \prd, 96,
  063007, \dodoi{10.1103/PhysRevD.96.063007}

\end{thebibliography}

\end{document}